\documentclass[iop,twocolappendix]{emulateapj}
\usepackage{times,epsfig} 
\usepackage{natbib}
\usepackage{amssymb,amsmath}
\usepackage{ctable}
\shorttitle{}
\shortauthors{De Boni et al.}

\begin{document}

\title{The mass accretion rate of galaxy clusters: a measurable quantity}

\author{C. De Boni\altaffilmark{1,2,3}}
\author{A.L. Serra\altaffilmark{4}} 
\author{A. Diaferio\altaffilmark{1,2}}
\author{C. Giocoli\altaffilmark{5,6,7,8}} 
\author{M. Baldi\altaffilmark{6,7,8}}

\affil{\altaffilmark{1}Dipartimento di Fisica, Universit\`a di Torino, via P. Giuria 1,
  I-10125 Torino, Italy; cristiano.deboni@ias.u-psud.fr} 
\affil{\altaffilmark{2}Istituto Nazionale di Fisica Nucleare (INFN), Sezione di Torino, via P. Giuria 1,
  I-10125 Torino, Italy}
\affil{\altaffilmark{3}Institut d'Astrophysique Spatiale, CNRS (UMR8617) Universit\'e Paris-Sud 11, B\^{a}t. 121, F-91405 Orsay, France}
\affil{\altaffilmark{4}Dipartimento di Fisica, Universit\`a degli Studi di Milano, via Celoria 16, I-20133 Milano,
Italy}
\affil{\altaffilmark{5}Aix Marseille Universit\'e, CNRS, LAM (Laboratoire d'Astrophysique de Marseille) UMR 7326, F-13388 Marseille, France}
\affil{\altaffilmark{6}Dipartimento di Fisica e Astronomia, Alma Mater Studiorum Universit\`a di Bologna, viale Berti Pichat 6/2, I-40127 Bologna, Italy}
\affil{\altaffilmark{7}Istituto Nazionale di Astrofisica (INAF), Osservatorio Astronomico di Bologna, via Ranzani 1, I-40127 Bologna, Italy}
\affil{\altaffilmark{8}Istituto Nazionale di Fisica Nucleare (INFN), Sezione di Bologna, viale Berti Pichat 6/2, I-40127 Bologna, Italy}

\begin{abstract}
We explore the possibility of measuring the mass accretion rate (MAR) 
of galaxy clusters from their mass profiles beyond the virial radius $R_{200}$. 
We derive the accretion rate from the mass of a spherical shell whose
inner radius is $2R_{200}$,  whose thickness changes with redshift, and
whose infall velocity is assumed to be equal to the mean infall velocity of the spherical shells    
of dark matter halos extracted  from $N$-body simulations.  
This approximation is rather crude in hierarchical clustering scenarios where 
both smooth accretion and aggregation of smaller dark matter halos contribute 
to the mass accretion of clusters. 
Nevertheless, in the redshift range $z=[0,2]$,
our prescription returns an average MAR within $20-40 \%$ of the average rate derived
from the merger trees of dark matter halos extracted from $N$-body simulations.
The MAR of galaxy clusters has been the topic of numerous 
detailed numerical and theoretical investigations, but so far it has remained 
inaccessible to measurements in the real universe.
Since the measurement of the mass profile of clusters beyond their virial radius can 
be performed with the caustic technique applied to dense redshift surveys of the
cluster outer regions,  
our result suggests that measuring 
the mean MAR of a sample of galaxy clusters
is actually feasible. We thus provide a new potential observational test of the cosmological
and structure formation models.
\end{abstract}


\section {Introduction}

In the current model of the formation of cosmic structure, where dark matter halos
form from the aggregation of smaller halos, 
the mass accretion of dark matter halos is a stochastic
process whose average behavior can be predicted with $N$-body simulations and semi-analytical models \citep{2002MNRAS.331...98V,2003MNRAS.339...12Z,2004MNRAS.349.1464S,2004MNRAS.350.1385S,2007MNRAS.376..977G,2009MNRAS.398.1858M,2009ApJ...707..354Z}. 
This process is generally investigated with the identification of the merger
trees of dark matter halos, enabling the study of the mass accretion history (MAH) and the mass accretion rate (MAR) as a function of redshift $z$ \citep{2002MNRAS.331...98V,2008MNRAS.386..577F,2009MNRAS.398.1858M,2010MNRAS.406.2267F,2012MNRAS.422..185G}.

Observationally, the exploration of the MAR of individual dark matter halos has only been attempted on the scales of galaxies with the galaxy--galaxy merger rate: one usually combines the number of observed pairs of close or disturbed galaxies with the theoretical merger probability and time scale \citep{2011ApJ...742..103L,2012ApJ...754...26J,2014MNRAS.445.1157C}. However, different investigations reach discrepant conclusions \citep{2011ApJ...742..103L} because the merger rate of dark matter halos is not identical to the merger rate of galaxies \citep{2008MNRAS.386..577F,2008MNRAS.384....2G,2013MNRAS.428.3121M} and the two rates are related by dissipative processes which are difficult to model \citep{2013MNRAS.430.1901H}. 

In contrast, the measurement of the MAR of galaxy clusters can in principle be
based on the estimated amount of mass in the cluster's surrounding regions, where we can safely
neglect the dissipative processes which affect the galaxy--galaxy merger rate. Nevertheless, no measurements
of the MAR of galaxy clusters have been attempted so far. This observational deficiency is due to the fact that in the large and less dense outer regions of clusters, galaxy members are difficult to distinguish from foreground and background galaxies; other probes, e.g. X-ray emission, are below the sensitivity of current instruments. In addition, the outer regions of clusters are not in dynamical equilibrium and therefore the usual mass estimation methods based on virial equilibrium are inappropriate. 
Considering this picture, one may think that the MAR predictions from $N$-body simulations are not capable of being tested. 

Here we take a more optimistic perspective and 
explore the possibility of estimating the MAR of galaxy clusters by measuring
the mass of a spherical shell surrounding the cluster. The thickness of this shell depends
on the assumed infall time, on the radius at which the infall happens and on the initial velocity of the falling mass. This translates into a change of the shell thickness with redshift. 
Albeit rather crude when compared with the stochastic aggregation of dark matter
halos in the hierarchical clustering formation model, this approach would provide
a method to estimate the MAR that depends on the cluster mass profile at radii larger than the virial radius. 

In theoretical investigations, the relation between the mass density profiles of galaxy clusters and their accretion history is known. 
For example, \cite{2013MNRAS.432.1103L} find that the inner part of a halo retains the information on how the halo has accreted its mass through a correlation between the mean inner density within the scale radius $r_s$ and the critical density of the universe at the time when the mass of the main progenitor is equal to $M(<r_s)$. \cite{2015MNRAS.450.1521C,2015MNRAS.450.1514C} confirm these findings and demonstrate that the MAH can be expressed with a general formula similar to the one originally proposed by \cite{2004ApJ...607..125T} for a $\Lambda$CDM model and widely studied in \cite{2009MNRAS.398.1858M} based on $N$-body simulations: the MAH has an exponential evolution with redshift in the high $z$ regime and follows a power law at low $z$ when the accelerated expansion of the universe freezes the growth of perturbations.

\cite{2014ApJ...789....1D} find that the steepness of the slope of the outer halo density profile increases with increasing MAR. In addition, the central concentration is anti-correlated with the MAR. \cite{2014JCAP...11..019A} note that the location where the steepening of the slope is observed corresponds to the radius associated to the splashback of the material that the halo has recently accreted.
\cite{2014MNRAS.445.1713V} show that the growth of the central potential precedes the assembly of mass and introduce a formula that can be used to compute the average MAH in any $\Lambda$CDM cosmology without running numerical simulations.

Here we derive a simple relation between the mass profile of a dark matter halo and its MAR 
derived from the halo merger tree extracted from $N$-body simulations.
This result is relevant because it implies that in principle 
we can estimate the MAR of galaxy clusters from the estimate of the mass 
profile in their outer regions. This measurement can be performed with the caustic technique \citep{1997ApJ...481..633D, 1999MNRAS.309..610D, 2011MNRAS.412..800S} which is not affected by the presence
of substructures, the non-equilibrium state of the cluster outer region, and the correlated large-scale structures along the line of sight \citep{1999MNRAS.309..610D,2011MNRAS.412..800S,2013ApJ...764...58G}. 
The caustic technique only requires a sufficiently dense galaxy redshift survey in the cluster field of view to return a mass estimate accurate to 20\% on average \citep{2011MNRAS.412..800S}. 

Estimating the MAR also requires the knowledge of the infall velocity of the shell. This quantity is inaccessible to observations and remains a free parameter; we adopt the mean infall velocity of
a shell surrounding dark matter halos in $N$-body simulations as illustrated below. Although it prevents us from estimating the MAR of individual clusters, this choice allows us to cope with the unmeasurable velocity of the falling shell. Consequently, the technique we propose here aims at estimating the mean accretion rate of a sample of galaxy clusters rather than the accretion rate of individual clusters.

We investigate the feasibility of our approach by considering dark matter halos at redshift $z<2$, with mass comparable to
the clusters in the CIRS and HeCS catalogs \citep{2006AJ....132.1275R,2013ApJ...767...15R} whose outer mass profiles have already
been measured with the caustic technique.
Our interest in investigating the growth of structures at nonlinear scales of galaxy clusters from an observational perspective, which is still relatively poorly explored \citep[{\it{e.g.}},][]{2013ApJ...776...91L}, perfectly complements most of the current efforts that focus on constraining the growth factor in the linear and mildly nonlinear regimes with large-scale redshift surveys and weak-lensing tomography ({\it{e.g.}}, Euclid, DES, eBOSS, DESI, PFS, LSST, and WFIRST). 

In Section \ref{model} we introduce our spherical infall prescription. In Section \ref{CoDECS} we discuss the properties of the CoDECS set of $N$-body simulations \citep{2012MNRAS.422.1028B}. 
We discuss our results in Section \ref{results}.

\section{The accretion recipe} \label{model}

Our spherical infall prescription assumes a shell of matter falling onto the enclosed halo.
We aim to quantify $\dot{M}=d M/ d t$, where $d M$ is the mass of a spherical shell of thickness $\Delta r$ and proper radii $R_{i}$
and $R_{i}+ \Delta r$ and $dt$ is the time it takes to fall to $R_{i}$. $R_{i}$ is the radius at which we consider the infall to happen. We choose $\Delta r= \delta_{s} R_{i}$, where $\delta_{s}$ is a free parameter.
By solving the equation of motion ${\rm{d}}^2 r/{\rm{d}}t^2 = a_{0}$ under the assumption of constant infall acceleration, $a_0 = -G M(<R_{i}) / (R_{i} + \delta_{s} R_{i}/2)^2$, and initial velocity $v_{i}$, we obtain the infall time $t_{\rm inf}$ from the equation $a_0t_{\rm inf}^2/2 + v_{i}t_{\rm inf}=-\delta_{s}R_{i}/2$, where we consider the shell to be accreted when the shell middle point, initially at  $\delta_{s} R_{i}/ 2$, reaches $R_{i}$; we obtain 

\begin{equation}
\begin{split}
t_{\rm inf}^2 GM(<R_{i}) & - t_{\rm inf} 2 R_{i}^2 (1+  \delta_{s}/2)^2 v_{i} + \\
& - R_{i}^3  \delta_{s} (1+  \delta_{s}/2)^2 = 0 \ .
\end{split}
\label{infall_time}
\end{equation}

We can now express the MAR obtained from our prescription as

\begin{equation}
\dot{M}=\frac{dM}{dt}\equiv \frac{M_{\rm shell}}{t_{\rm inf}} \, .
\label{MAR_recipe}
\end{equation}

This recipe has three input parameters: the scale $R_{i}$ that defines the infall radius, the initial velocity $v_{i}$ and the thickness $\delta_{s}$ of the shell.
It is more convenient to use the infall time as an input parameter and to derive the shell thickness $\delta_{s}$. In this case, Equation (\ref{infall_time}) reads:

\begin{equation}
\begin{split}
\delta_{s}^3 \frac{R_{i}^3}{4} & + \delta_{s}^2 \left( R_{i}^3 + \frac{R_{i}^2}{2} v_{i} t_{\rm inf} \right) + \\
& + \delta_{s} \left( R_{i}^3 + 2 R_{i}^2 v_{i} t_{\rm inf} \right) = \\
& GM(<R_{i}) t_{\rm inf}^2 - 2R_{i}^2 v_{i} t_{\rm inf} \ .
\end{split}
\label{shell_thickness}
\end{equation}

Therefore, by choosing $t_{\rm inf}$, we derive the thickness  $\delta_{s} R_{i}$ of the shell 
centered on the cluster and we can estimate the MAR of a galaxy cluster from its mass profile
by using Equation (\ref{MAR_recipe}).  

However, accretion is a stochastic process, and we need to verify that our simple approach is capable of correctly estimating  the actual MAR. 
Below, we compare 
the MAR estimated with our recipe with the MAR of dark matter halos derived from 
their halo merger trees obtained in an $N$-body simulation.

\section{CoDECS simulations} \label{CoDECS}

CoDECS (Coupled Dark Energy Cosmological Simulations) is a suite of $N$-body simulations in different cosmological
models \citep[see][for further details]{2012MNRAS.422.1028B}. Here we use the L-CoDECS
simulation of a $\Lambda$CDM model, a collisionless $N$-body simulation of a flat universe, with the following cosmological
parameters consistent with the {\it{WMAP7}} data \citep{2011ApJS..192...18K}: cosmological matter density $\Omega_{m0}=0.226$, cosmological constant $\Omega_{\Lambda 0}=0.729$, baryonic mass density $\Omega_{b0}=0.0451$, Hubble constant $h=0.703$, power spectrum normalization $\sigma_{8}=0.809$, and power spectrum index $n_{s}=0.966$. The simulated box has a comoving volume of $(1 \ {\rm{Gpc}} \ h^{-1})^{3}$ containing $(1024)^{3}$ dark matter (CDM) particles and the same amount of baryonic particles. The mass resolution is $m_{\rm DM} = 5.84 \times 10^{10} \ {\rm{M_{\odot}}} \ h^{-1}$ for CDM particles and $m_{b} = 1.17 \times 10^{10} \ {\rm{M_{\odot}}} \ h^{-1}$ for baryonic particles. No hydrodynamics are included in the simulation. Baryonic particles are only included to account for the different forces acting on baryonic matter in the coupled quintessence models.

\subsection{CoDECS Merger Trees} \label{tree}

Dark matter halos are identified at a given time with a Friends-of-Friends (FoF) algorithm with linking length $b=0.2$ times the mean interparticle separation. Halos identified with the FoF algorithm are called FoF halos. Halo substructures are identified with the SUBFIND algorithm \citep{2001MNRAS.328..726S}. We call SUBFIND halo the halo substructure that contributes the most particles to the FoF halo. To each FoF halo, we assign the radius $R_{200}$\footnote[9]{$R_{\Delta}$ is the radius of the sphere of mass $M_{\Delta}$ centered on a local minimum of the gravitational potential with average density $\Delta$ times the critical density of the universe, $\rho_{c} \equiv 3H^2(z)/8 \pi G$, where $H$ is the Hubble parameter.} and the corresponding
enclosed mass $M_{200}$ of its SUBFIND halo.

To derive the MAH of each halo, we use the merger trees provided in the CoDECS public database (\url{www.marcobaldi.it/CoDECS}). Each FoF halo at a given time $t_i$ has a SUBFIND halo.
We trace the main branch of this SUBFIND halo by searching for its main progenitor at each previous time step. 
The main progenitor of a SUBFIND halo at time $t_i$ is the SUBFIND halo at time $t_{i-1}<t_i$
which contains the largest number of particles that will end up in the
same FoF halo of the SUBFIND halo at time $t_i$. To each halo we associate 
the mass $M_{200}$ of its SUBFIND halo and thus we derive the MAH as $M_{200}(z)$ of the SUBFIND halos 
along the main branch. 

The definitions of the SUBFIND halo and the main progenitor are not unique in the literature. The SUBFIND halo can also be defined as the substructure with the largest $M_{200}$ in the halo, rather than the substructure with the largest number of
particles we adopt here. The usual definition of the main progenitor also is slightly different from ours: it can be defined either as the subhalo at $t_{i-1}$ that donates the largest number of particles to the SUBFIND halo at $t_i$ or the most massive progenitor of the SUBFIND halo. In general, the differences between our MAHs and the MAHs 
obtained with the more common definitions of SUBFIND halo and main progenitor are negligible. However, our 
definitions guarantee that we always trace the branch of the merger tree with the most massive halos.

\subsection{Cluster profiles} \label{profiles}

We plan to apply our recipe for the estimate of the MAR to redshift surveys of galaxy clusters 
by estimating their mass profile with the caustic method \citep{1997ApJ...481..633D, 1999MNRAS.309..610D, 2011MNRAS.412..800S}. The caustic method returns mass profiles that are affected by 20-50\% uncertainties when applied to clusters with $M_{200}$ of
$10^{14} \ {\rm M_{\odot}} \ h^{-1}$ or larger. When applied to less massive clusters, the systematic
errors introduced by the caustic technique can become substantially larger \citep{2010AJ....139..580R, 2011MNRAS.412..800S}.
Therefore, with this observational perspective in mind here we concentrate on halos in the $M_{200}$ mass range $10^{14}-10^{15} \ {\rm{M_{\odot}}} \ h^{-1}$, which corresponds to the most common mass of the clusters in catalogs like CIRS \citep{2006AJ....132.1275R} and HeCS \citep{2013ApJ...767...15R}. We thus define two mass bins at $z=0$ and follow the evolution with redshift of the clusters assigned to these bins.
We select halos with median $M_{200}= 10^{14}$ and $10^{15} \ {\rm{M_{\odot}}} \ h^{-1}$ at $z=0$. The low-mass bin contains $2000$ objects at $z=0$, while the high-mass bin is limited to $50$
objects: there are $36$ clusters with $M_{200} > 10^{15} \ {\rm{M_{\odot}}} \ h^{-1}$ in the simulated box, but we removed the $11$ most massive clusters in order to have a mean mass similar to the median mass in the bin. Table \ref{mass_2bins} lists the mean, standard deviation, median, $68 \%$, and $90 \%$ percentiles of each mass bin at $z=0$.

\begin{table*}
\begin{center}
\caption{Mean, Median, Standard Deviation, and Percentile Ranges of $M_{200}$ and Number of Halos in the Two Mass Bins at $z=0$}
\begin{tabular}{lccccc}
\hline
\hline
\\
Mean & $\sigma$ & Median & $68\%$ & $90\%$ & Number of Halos \\
\\
\hline
\\
\multicolumn{5}{c}{$M_{200} \ [{\rm{M_{\odot}}} \ h^{-1}]$} & \\
\\
\cline{1-5}
\\
$1.00 \times 10^{14}$ & $2.90 \times 10^{12}$ & $1.00 \times 10^{14}$ & $(0.97-1.04) \times 10^{14}$ & $(0.96-1.05) \times 10^{14}$ & $2000$ \\
\\
\hline
\\
$1.04 \times 10^{15}$ & $1.26 \times 10^{14}$ & $1.00 \times 10^{15}$ & $(0.91-1.19) \times 10^{15}$ & $(0.88-1.24) \times 10^{15}$ & $50$\\
\\
\hline
\end{tabular}
\label{mass_2bins}
\end{center}
\end{table*}

\begin{figure*}
\hbox{
 \epsfig{figure=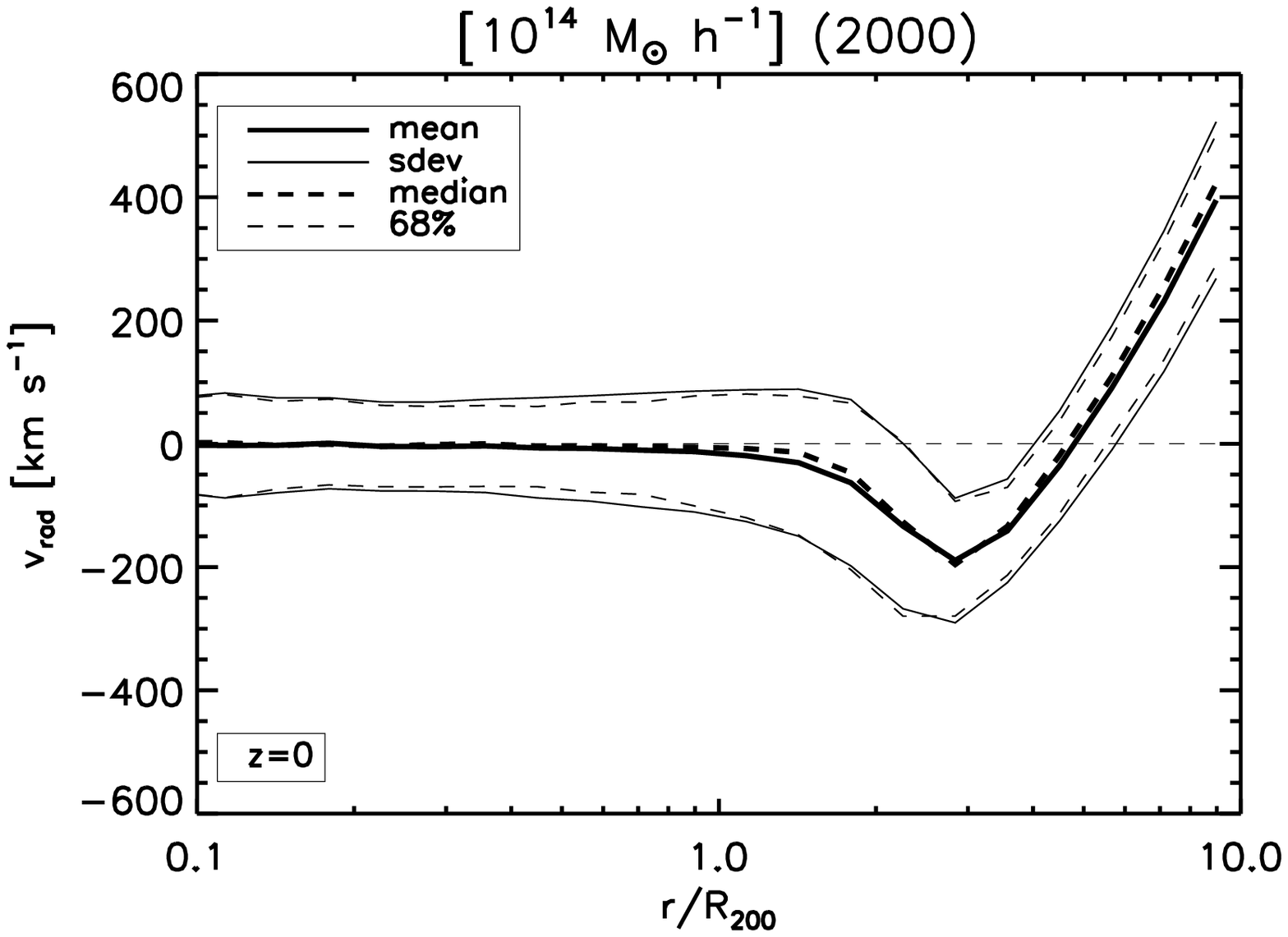,width=0.45\textwidth}
 \epsfig{figure=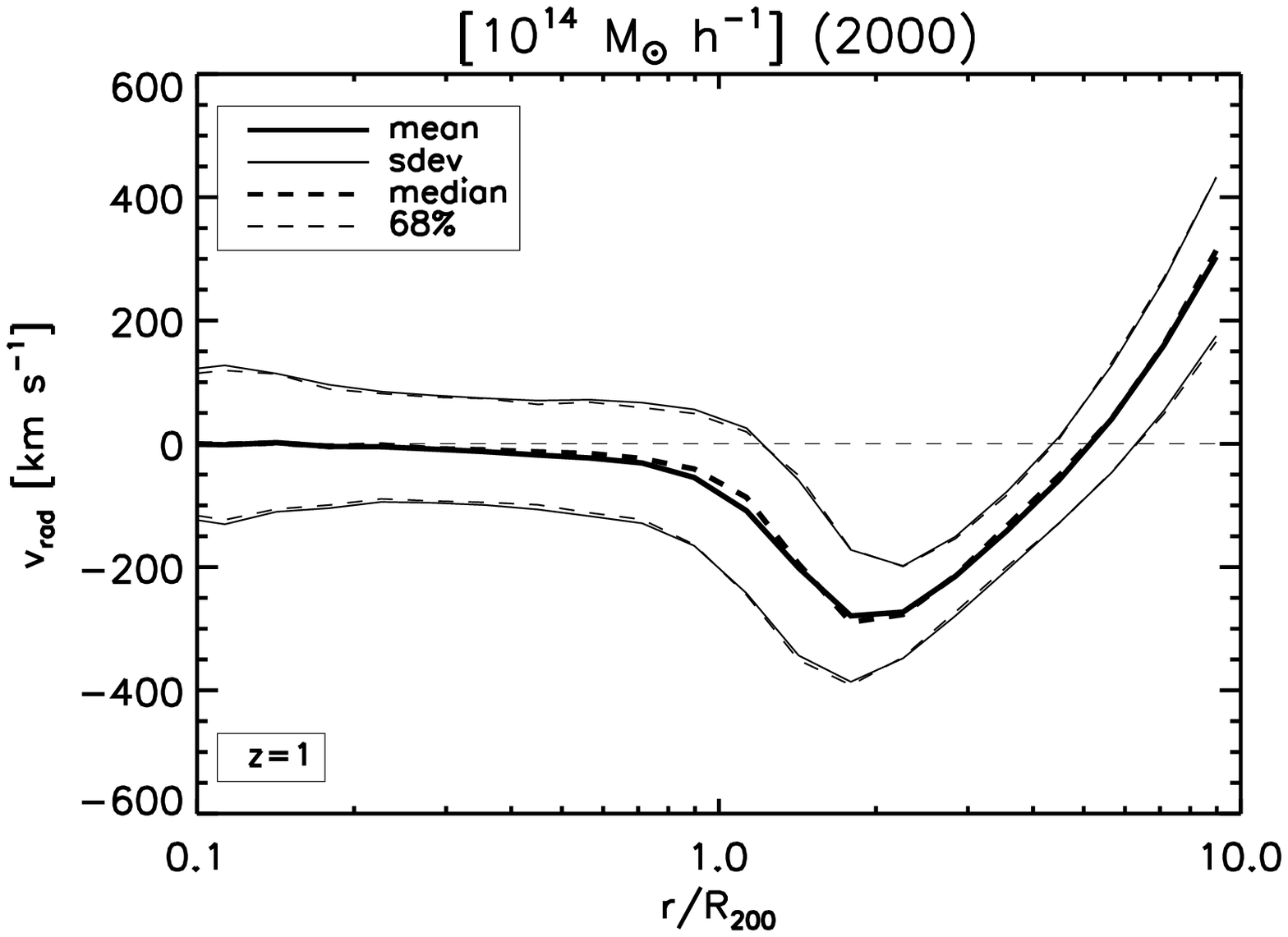,width=0.45\textwidth}
}
\hbox{
}
\caption{Radial velocity profile for clusters in the $10^{14} \ {\rm{M_{\odot}}} \ h^{-1}$ mass bin at $z=0$ (left panel) and $z=1$ (right panel). The thick-solid and thick-dashed lines indicate the mean and median profiles, respectively. The thin-solid and thin-dashed lines indicate the standard deviation and the $68\%$ of the distribution, respectively.}
\label{velocity_profile_14}
\end{figure*}

\begin{figure*}
\hbox{
 \epsfig{figure=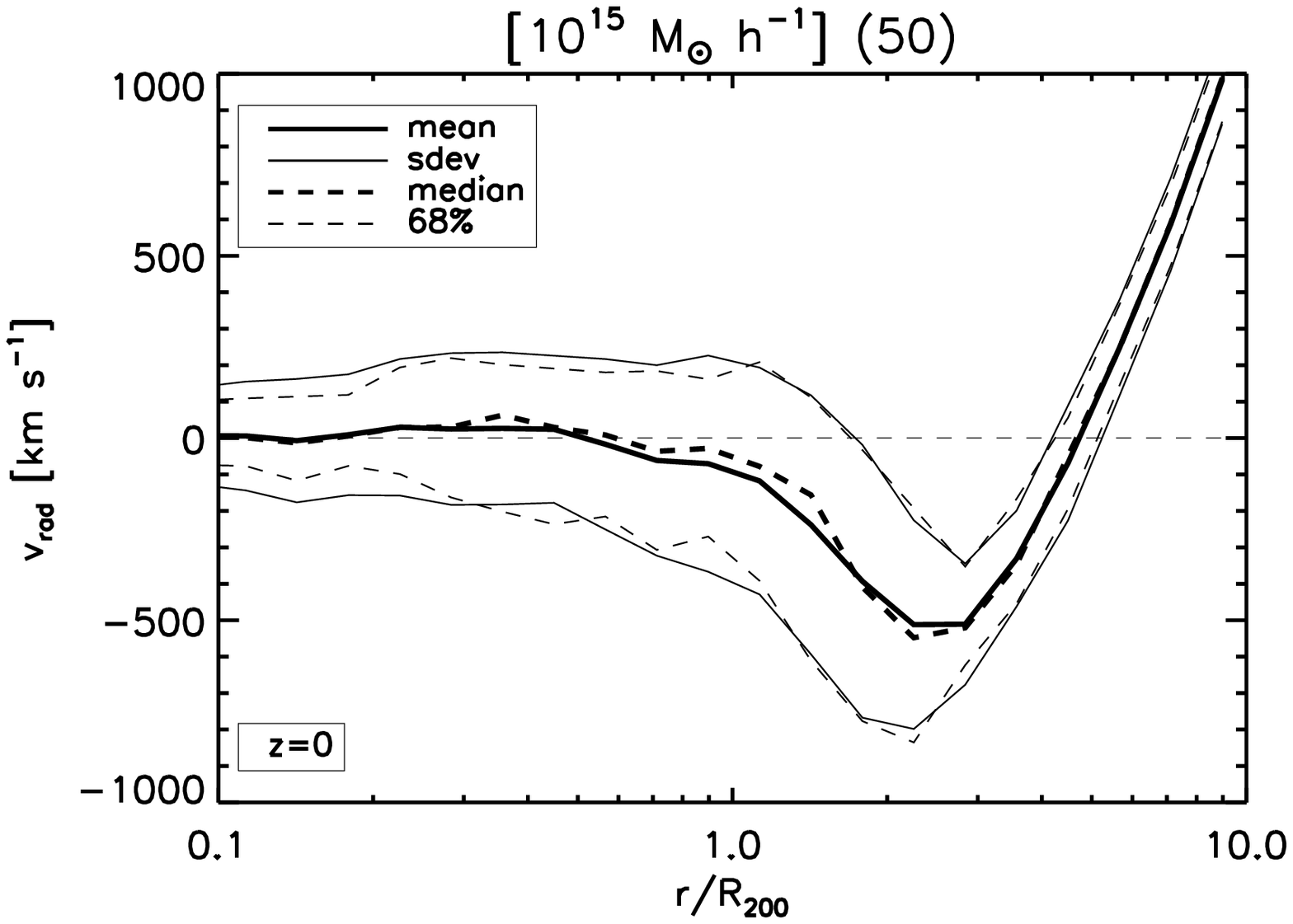,width=0.45\textwidth}
 \epsfig{figure=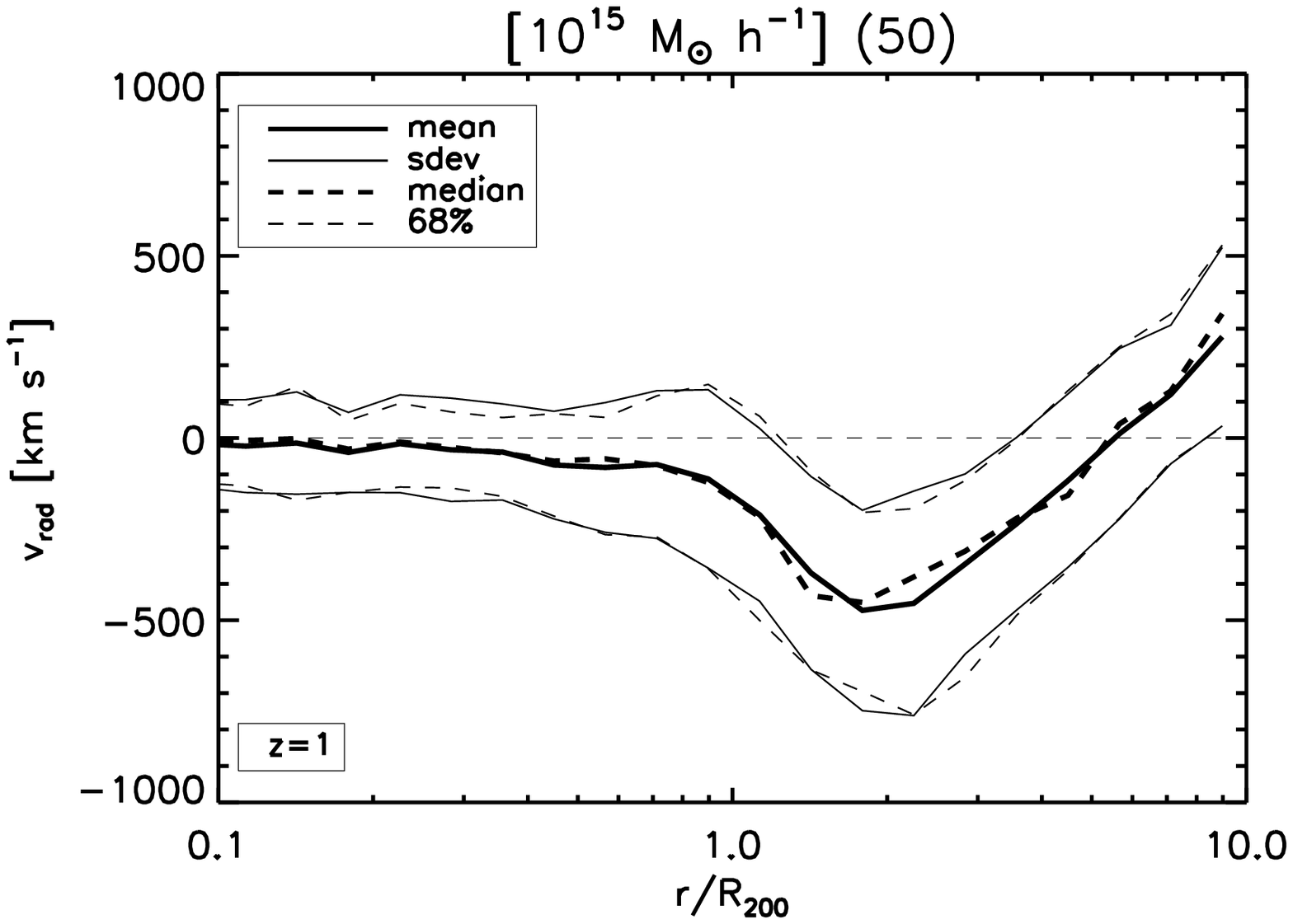,width=0.45\textwidth}
}
\hbox{
}
\caption{Same as Figure \ref{velocity_profile_14} except for the $10^{15} \ {\rm{M_{\odot}}} \ h^{-1}$ mass bin.}
\label{velocity_profile_15}
\end{figure*}

For each halo in the two mass bins and for each progenitor at higher redshift (as defined in Section \ref{tree}), we evaluate the mass profile and the profile of the radial velocity $v_{\rm rad}$ up to $10R_{200}$.
We show the radial velocity profile in the two bins at different redshifts in Figures \ref{velocity_profile_14} and \ref{velocity_profile_15}.

With the radial velocity profile we can identify three regions: an internal region with $v_{\rm rad} \simeq 0$, where matter is orbiting around the center of the cluster; an infall region, where $v_{\rm rad}$ becomes negative and indicates an actual infall of matter toward the center of the cluster; and
a Hubble region at very large radii, 
where $v_{\rm rad}$ becomes positive and the Hubble flow dominates.
Broadly speaking, the infall radius $R_{\rm inf}$, i.e., the radius where the minimum of 
$v_{\rm rad}$ occurs, is between $2R_{200}$ and $3R_{200}$, independent of mass and redshift.

\cite{2015ApJ...810...36M} suggest the use of the splashback radius as the physical halo boundary because it separates the infall region from the region where the matter has already been accreted. The splashback radius is defined as the outermost radius reached by accreted material in its first orbit around the cluster center. Its exact location depends both on redshift and MAR, but it is in general larger than $R_{200}$. Noticeably, they also show that the splashback radius is not affected by the evolution of the critical density of the universe, unlike the usual $R_{200}$, whose definition is based on the average overdensity of the halo at a given redshift. By adopting $R_{200}$, part
of the evolution of the halo properties with redshift simply is a consequence of the evolution
of the critical density of the universe; this effect generates a so-called pseudo-evolution of the halo 
properties \citep[see][]{2013ApJ...766...25D}.
This pseudo-evolution substantially disappears when we adopt the splashback radius. This use of this radius might
thus be more preferable in the investigations of the redshift evolution of the halo properties.

Our results in Figures \ref{velocity_profile_14} and \ref{velocity_profile_15} are consistent with the results of \cite{2015ApJ...810...36M}. For massive objects at $z=0$, the splashback radius $R_{\rm sp}$ is close to $2R_{200}$ and the infall radius is $R_{\rm inf} \sim 1.4 R_{\rm sp}$.
Given that $M(<R_{\rm sp})$ and $M(<R_{\rm inf})$ are not affected by the pseudo-evolution mentioned above, we use $R_{i}=2R_{200} \sim R_{\rm sp}$ as the radius at which we consider the infall to happen in our spherical infall prescription and $v_{i}=v_{\rm shell} \sim v_{\rm rad}(R_{\rm inf})$ (see Equation (\ref{shell_thickness})).

\begin{figure*}
\hbox{
 \epsfig{figure=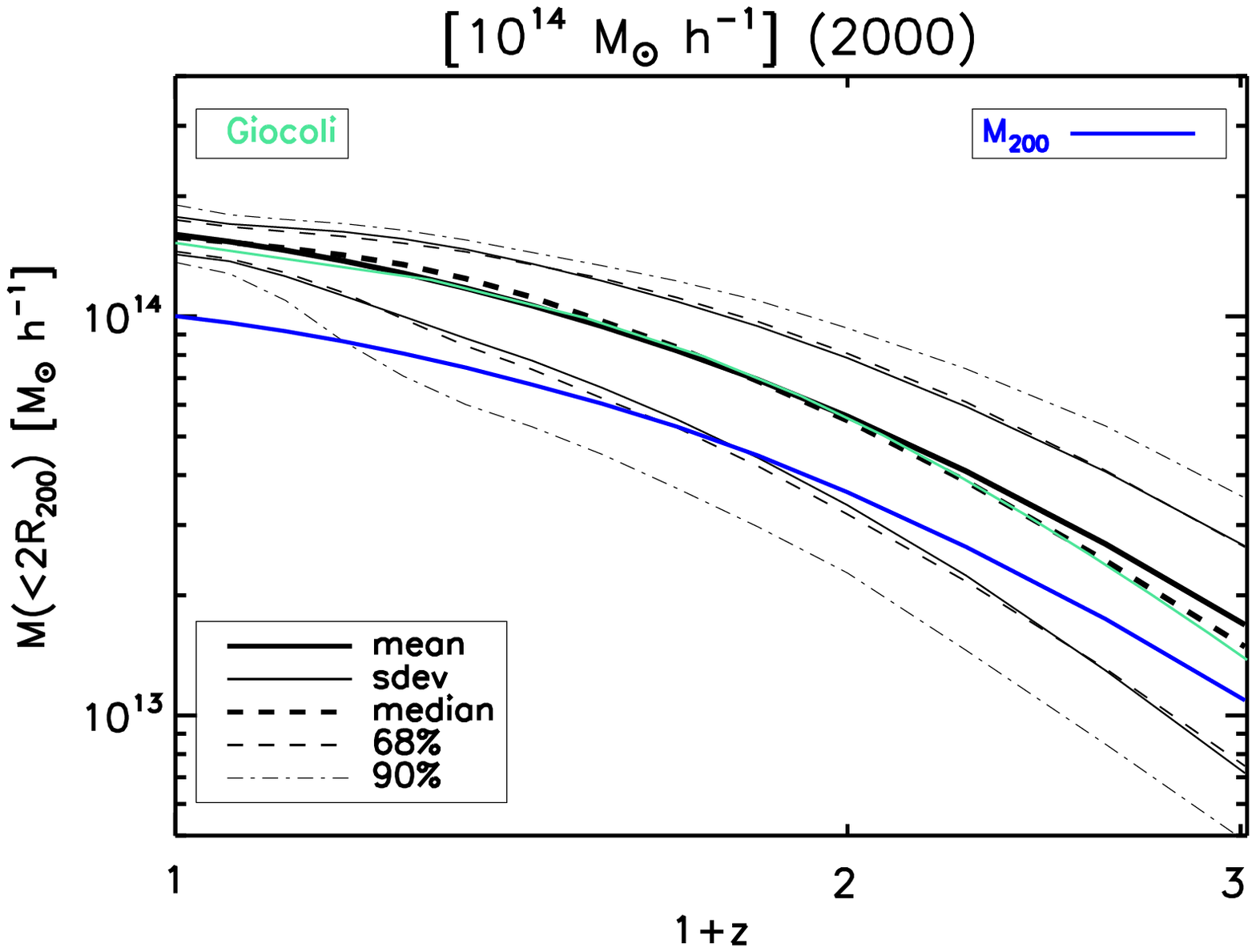,width=0.45\textwidth}
 \epsfig{figure=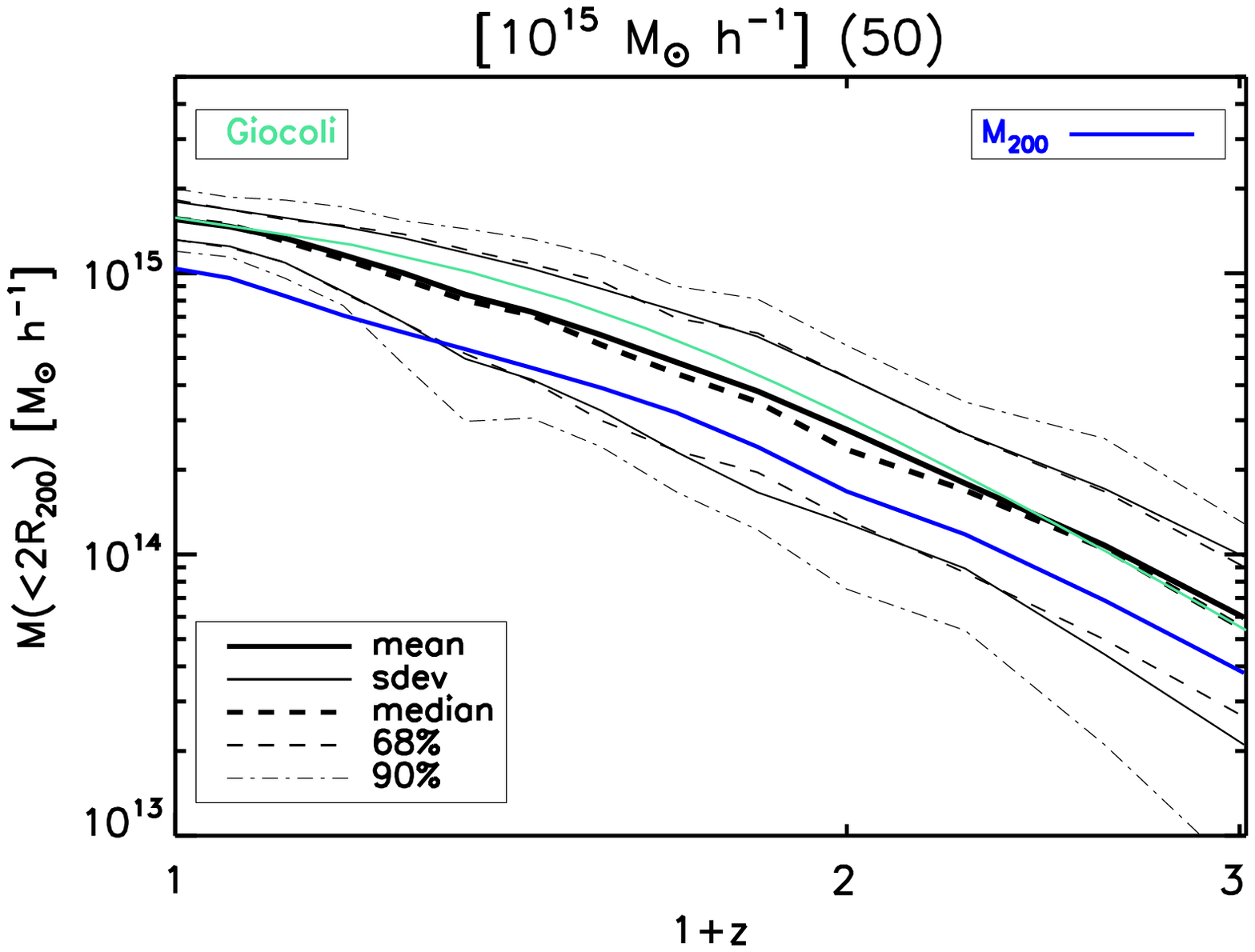,width=0.45\textwidth}
}
\hbox{
}
\caption{MAH at $2R_{200}$ for the $10^{14} \ {\rm{M_{\odot}}} \ h^{-1}$ (left panel) and $10^{15} \ {\rm{M_{\odot}}} \ h^{-1}$ (right panel) mass bins. The thick-solid and thick-dashed lines indicate the mean and median MAH, respectively. The thin-solid and thin-dashed lines indicate the standard deviation and the $68\%$ of the distribution, respectively. The thin-dotted-dashed line shows indicates $90\%$ of the distribution. The blue curve indicates the mean MAH for $M_{200}$. The green curve indicates the model by
\cite{2013MNRAS.434.2982G} rescaled to $2R_{200}$.}
\label{MAH_2R200}
\end{figure*}

For each halo at $z=0$ in the two mass bins we build the MAH at $2R_{200}$ (Figure \ref{MAH_2R200}). 
In Table \ref{mass_bins_2R200} we list the mean, standard deviation, median, $68 \%$, and $90 \%$ percentiles of $M(<2R_{200})$ at $z=0$ for the two mass bins. In Figure \ref{MAH_2R200} we also show the MAH model by
\cite{2013MNRAS.434.2982G} (see Appendix) rescaled to $2R_{200}$. We obtain this rescaling by extending the mass density profile to $2R_{200}$ using the NFW functional and adopting the redshift-dependent relation between the concentration and mass of \cite{2003ApJ...597L...9Z} modified for $M_{200}$ by \cite{2013MNRAS.434.2982G}. For a direct
comparison with previous results in the literature we also estimate the MAH at the
standard $R_{200}$. We report this analysis and the comparison with previous work in the Appendix.

\begin{table*}
\begin{center}
\caption{Mean, Median, Standard Deviation, and Percentile Ranges of $M(<2R_{200})$ in Each Mass Bin at $z=0$}
\begin{tabular}{lcccc}
\hline
\hline
\\
Mean & $\sigma$ & Median & $68\%$ & $90\%$ \\
\\
\hline
\\
\multicolumn{5}{c}{$M(<2R_{200}) \ [{\rm{M_{\odot}}} \ h^{-1}]$} \\
\\
\hline
\\
$1.60 \times 10^{14}$ & $1.73 \times 10^{13}$ & $1.58 \times 10^{14}$ & $(1.45-1.74) \times 10^{14}$ & $(1.37-1.90) \times 10^{14}$ \\
\\
\hline
\\
$1.55 \times 10^{15}$ & $2.41 \times 10^{14}$ & $1.56 \times 10^{15}$ & $(1.32-1.82) \times 10^{15}$ & $(1.19-2.00) \times 10^{15}$ \\
\\
\hline
\end{tabular}
\label{mass_bins_2R200}
\end{center}
\end{table*}

We find the ratio $M(<2R_{200})/M_{200}$ to be $\sim 1.6$, similar to the result of \cite{2015ApJ...810...36M} for $M_{\rm sp}/M_{200}$ in massive objects at $z=0$. In real observations measuring the infall radius of a cluster, namely where the infall velocity reaches its minimum is currently unfeasible so we keep fixed $R_{i}=2R_{200}$ for all masses at all redshifts. This choice clearly is an oversimplification, but Figures \ref{velocity_profile_14} and \ref{velocity_profile_15} indicate that this assumption is reasonable. Indeed $R_{\rm inf}$ lies within $2R_{200}$ and $3R_{200}$ for a wide range of masses and redshifts.

\section{Our accretion recipe versus the merger trees} \label{results}

When we observe a cluster we look at a particular instant of its evolution and we can
trace the evolution of its mass neither backward nor forward in time. Therefore, the MAH is
not a measurable quantity. In contrast, the MAR can in principle be 
estimated as the ratio of the mass that is being accreted by the cluster at a given time and its infall
time as described by our spherical infall prescription in Section \ref{model}, Equation (\ref{MAR_recipe}). This estimate depends on the external mass
profile and on the mean infall velocity. The mass profile is measurable with the caustic method, whereas
the initial velocity $v_{i}$ (see Equation (\ref{shell_thickness})) remains a free parameter which can be inferred from $N$-body simulations.

As already stated in Section \ref{profiles}, we choose $R_{i}=2R_{200} \sim R_{\rm sp}$ as the radius where the infall occurs. As initial velocity $v_{i}$ for the infall,  for each redshift in each mass bin we adopt the mean velocity in our first radial bin $[2- 2.5] R_{200}\sim R_{\rm inf} $ (see Figures \ref{velocity_profile_14} and \ref{velocity_profile_15}); this velocity ranges from $-200$ to $-250 \ {\rm{km \ s^{-1}}}$. We call this velocity $v_{\rm shell}$. The value of $v_{\rm rad}$ in the radial bin $[2.5- 3] R_{200}$ roughly remains in the same range and adopting this radial bin instead of the first bin has negligible effects on the final results. Although our prescription for the MAR estimate clearly depends on the choice of both $R_{i}$ and $v_{\rm shell}$, using the mean infall velocity instead of the infall velocity of each halo enables the design of a feasible procedure for the observational estimate of the MAR of clusters.

The spread of the distribution of  $v_{\rm shell}$, estimated as $\sigma = \sigma_{v_{i}}/\sqrt{N}$, where $N$ is the number of halos used to evaluate $v_{\rm shell}$, is smaller than the spread of the distribution of the individual $v_{i}$'s, $\sigma_{v_{i}}$, shown in Figures \ref{velocity_profile_14} and \ref{velocity_profile_15}. For the $10^{14} {\rm M}_\odot h^{-1}$ mass bin, the $1 \sigma$ relative spread of $v_{\rm shell}$ is smaller than $2 \%$ and propagates into a relative spread on the MAR of $5  \%$ at most, which is well within the $68 \%$ percentiles of the MAR distribution obtained with $ v_{\rm shell}$. Similarly, for the $10^{15} {\rm M}_\odot h^{-1}$ mass bin, where $N$ drops from $2000$ to $50$, the $1 \sigma$ relative spread of $v_{\rm shell}$ is $\sim 14 \%$, implying a $\sim 35\%$ relative spread on the MAR. 
Incidentally, even by using the $1 \sigma$ spread of the distribution of the individual $v_{i}$'s, the resulting MAR's are within the $68 \%$ percentiles of the distribution of the MAR's obtained with $ v_{\rm shell}$ for both mass bins. 
The different spread of $v_{\rm shell}$ in the two mass shells deriving from the different size of the two cluster samples indicates that the expected uncertainty on the MAR estimated with real data will vary substantially depending on the number of observed clusters, as one can expect. We plan to fully quantify this uncertainty in future work by estimating the MAR of simulated clusters from mock redshift surveys.

For the infall time, which is the last parameter of the model, we choose $t_{\rm inf}=10^9 \ {\rm{yr}}$. This value is suggested by the redshift-independent $\sim 1 \ {\rm{Gyr}}$ time step of the snapshots of the simulation used to estimate the MAR from the merger trees; it also has the advantage of being similar to the dynamical time, simply defined as $t_{\rm dyn} \sim R/\sigma$, for the clusters of our analysis. In fact, for the $10^{14}-10^{15} \ {\rm{M_{\odot}}} \ h^{-1}$ clusters at $z=0$ we consider here, $R\sim 1 \ {\rm{Mpc}}$ and $\sigma\sim 1000 \ {\rm{km \ s^{-1}}}$, and $t_{\rm dyn} \sim10^9 \ {\rm{yr}}$.
For the progenitors of these clusters at higher redshifts, which have masses at most a factor $10$ smaller, the velocity dispersion $\sigma$ is $10^{1/2} \sim 3$ times smaller and the virial radius $R$ is smaller by roughly a similar factor. Therefore, $t_{\rm dyn}$ remains basically constant and equal to $10^9 \ {\rm{yr}}$. Finally, this equality between $t_{\rm dyn}$ and the snapshot time interval prevents us from assuming $t_{\rm inf}$ very different from $1 \ {\rm{Gyr}}$; if $t_{\rm inf}$ departs too much from the snapshot time interval, the comparison of our estimated MAR's with the MAR's extracted from the merger trees would be inappropriate. We therefore investigate the dependence of our results on $t_{\rm inf}$ within $\sim 20 \%$ of $10^9 \ {\rm{yr}}$ and find that our results remain unaffected.

\begin{figure*}
\hbox{
 \epsfig{figure=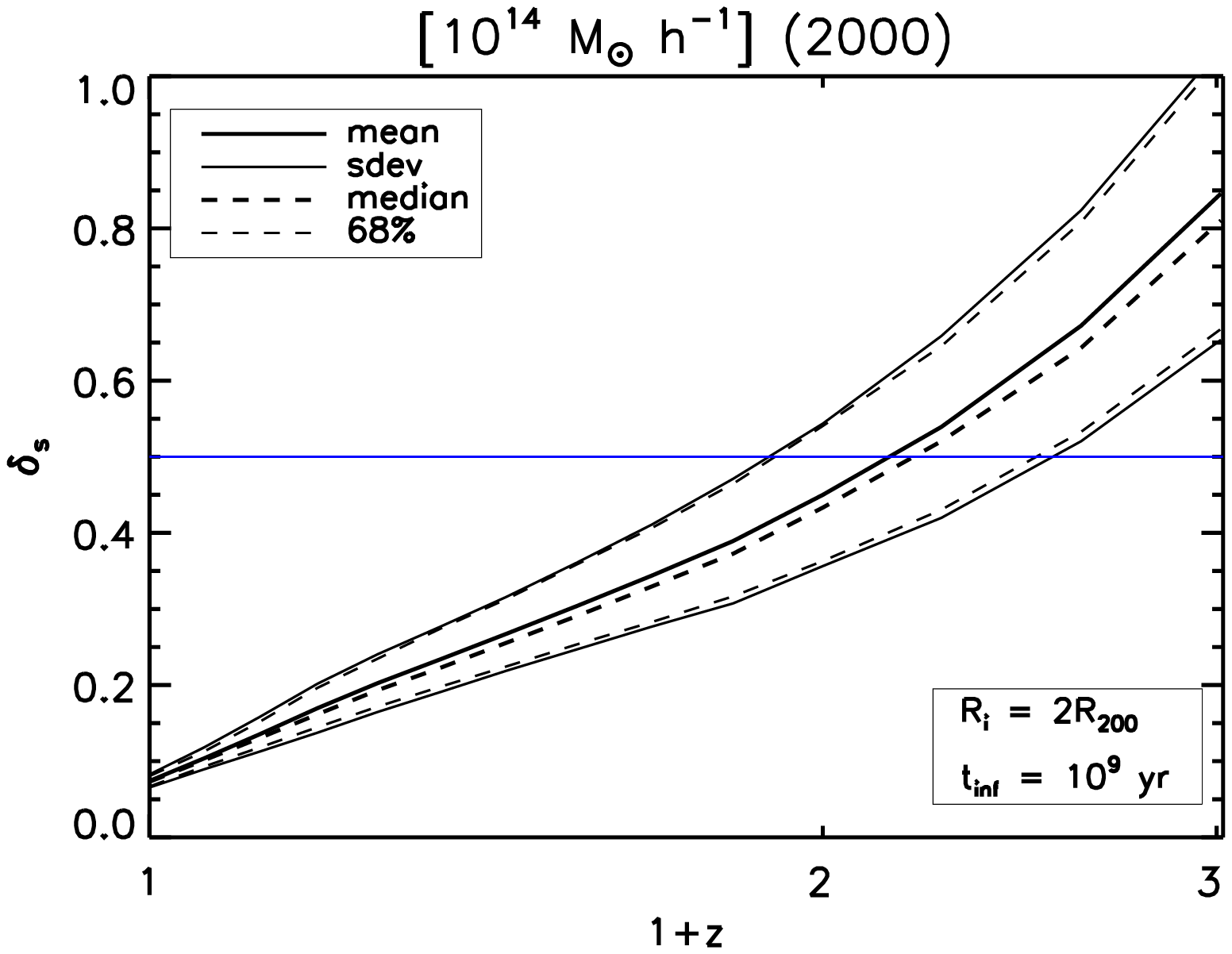,width=0.45\textwidth}
 \epsfig{figure=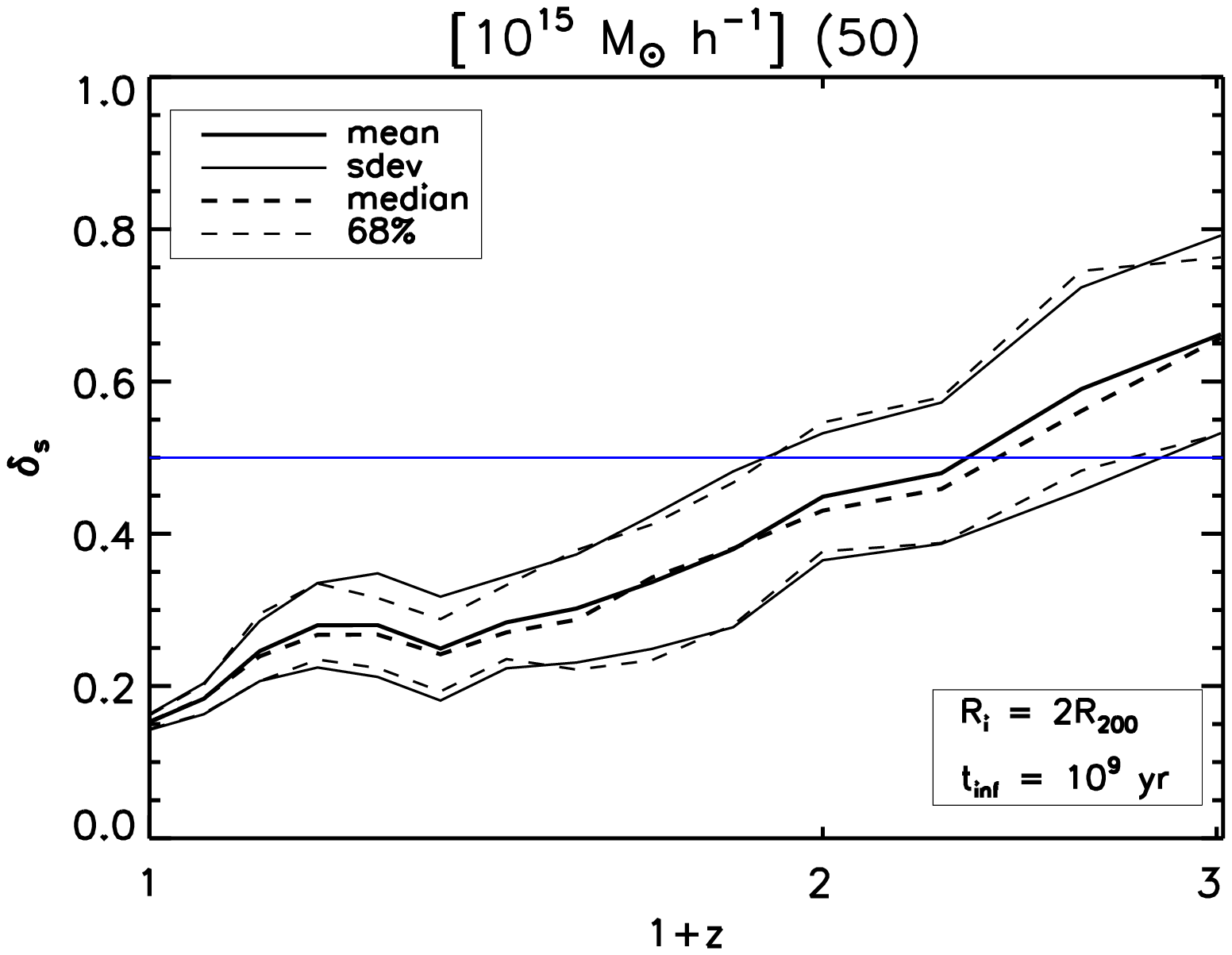,width=0.45\textwidth}
}
\hbox{
}
\caption{Redshift evolution of the shell thickness $\delta_{s}$ for the $10^{14} \ {\rm{M_{\odot}}} \ h^{-1}$ (left panel) and $10^{15} \ {\rm{M_{\odot}}} \ h^{-1}$ (right panel). The thick-solid and thick-dashed lines indicate the mean and median $\delta_{s}$, respectively. The thin-solid and thin-dashed lines indicate the standard deviation and the $68\%$ of the distribution, respectively. The solid blue line marks the value $\delta_{s}=0.5$ for which the external radius of the shell is equal to $3R_{200}$.}
\label{shell_thick_vs_z}
\end{figure*}

Once $R_{i}$, $v_{\rm shell}$ and $t_{\rm inf}$ are specified, the model is completely determined by Equation (\ref{shell_thickness}). For each halo in the two mass bins and for each progenitor at higher redshift, we evaluate the thickness $\delta_{s}$ of the infalling shell and its mass. We show the evolution with redshift of the shell thickness $\delta_{s}$ in Figure \ref{shell_thick_vs_z}. The shell thickness increases with increasing redshift, and the intrinsic scatter of the distribution also enlarges. This fact reflects the individual evolution of $M(<2R_{200})$, as shown in Figure \ref{MAH_2R200}. The solid blue line in Figure \ref{shell_thick_vs_z} marks the value $\delta_{s}=0.5$ for which the external radius of the shell is equal to $3R_{200}$, close to the cluster turnaround radius. This value of $\delta_s$ is reached between redshift $1$ and $1.5$, depending on the mass of the cluster. 

The accretion onto a cluster is a highly anisotropic process; nevertheless, we are confident that, given the thickness of the shell, we are taking into account almost all of the mass that is actually falling in the time interval $t_{\rm inf}$.

\begin{figure*}
\hbox{
 \epsfig{figure=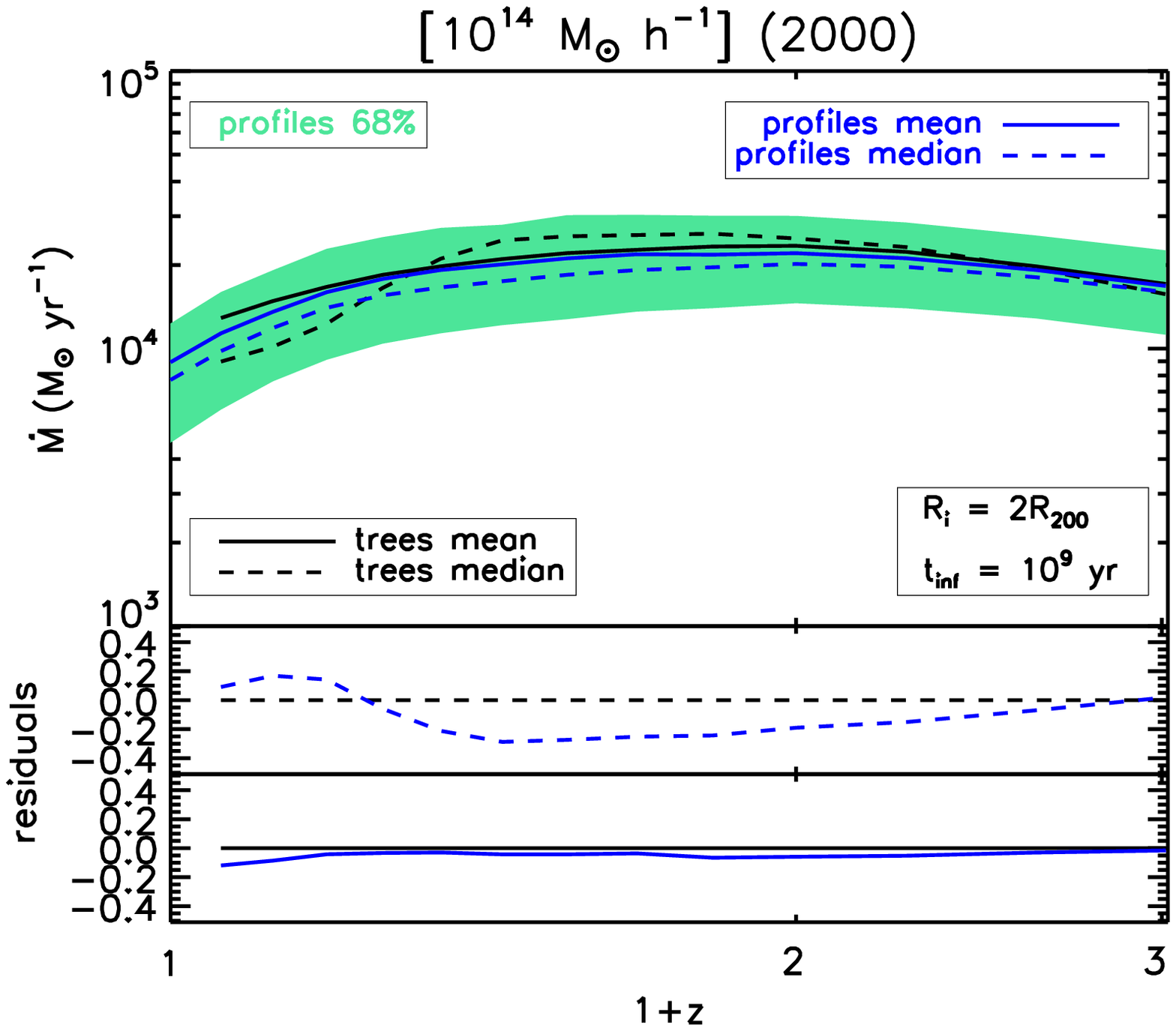,width=0.45\textwidth}
 \epsfig{figure=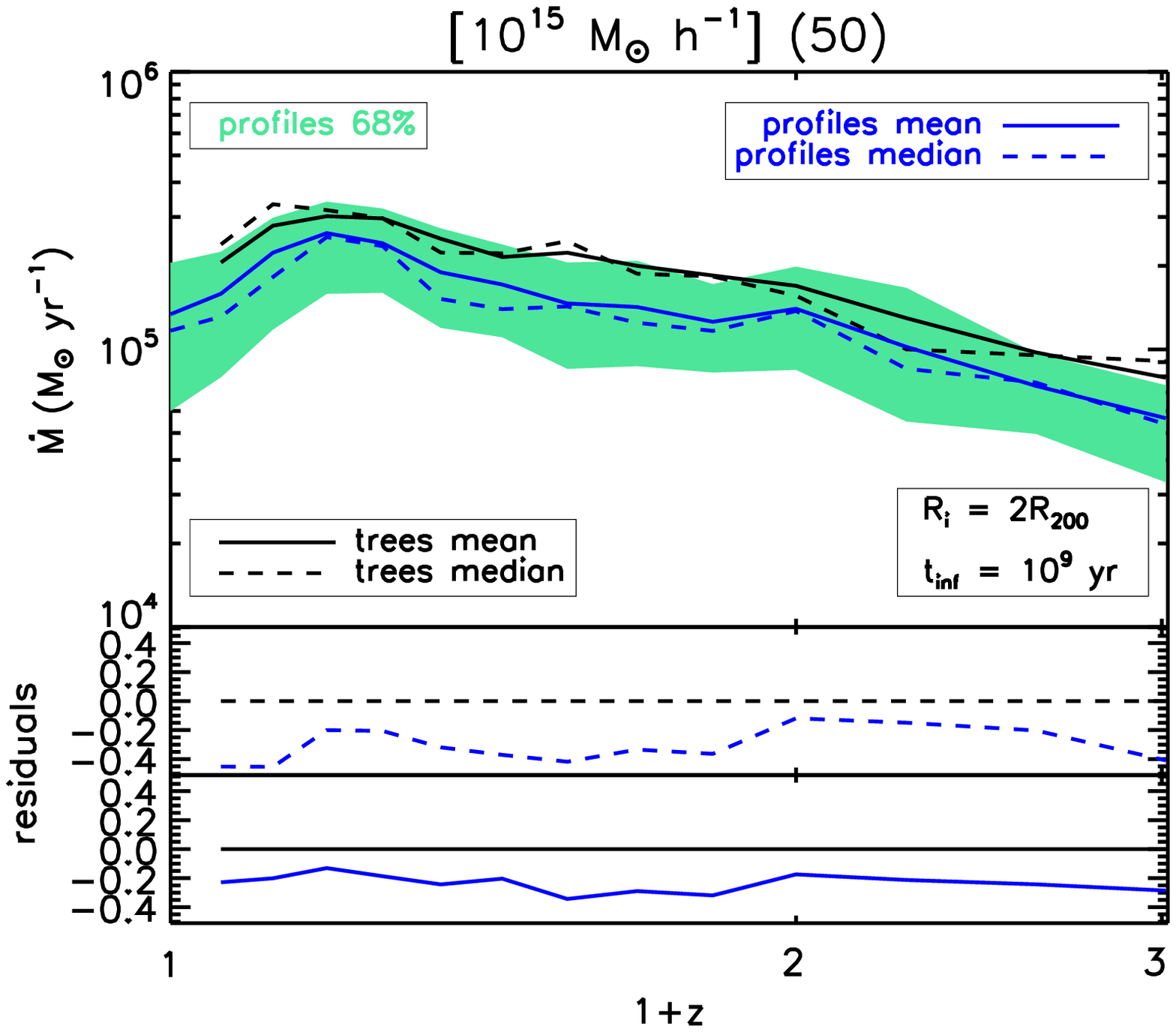,width=0.45\textwidth}
}
\hbox{
}
\hbox{
}
\hbox{
}
\caption{Results of our spherical infall model for the $10^{14} \ {\rm{M_{\odot}}} \ h^{-1}$ (left panel) and $10^{15} \ {\rm{M_{\odot}}} \ h^{-1}$ (right panel) mass bins and comparison with the MAR from merger trees. 
The blue solid and dashed lines are the mean and median MAR from Equation (\ref{MAR_recipe}). The green area indicates the $68\%$ of this MAR distribution. The mean and median MAR from the merger trees are indicated by the black solid and dashed lines. Residuals from the median and the mean values are shown in the insets at the bottom of each panel.}
\label{spherical_infall}
\end{figure*}

Figure \ref{spherical_infall} shows the MAR of all the clusters in the two mass bins estimated with Equation (\ref{MAR_recipe}). It also shows the MAR derived from the merger trees of the halos.

We see that the mean and median results from the merger trees lie within the region defined by the $68 \%$ percentile range of the distribution of the MAR's obtained from Equation (\ref{MAR_recipe}) for both mass bins. For the $10^{15} \ {\rm{M_{\odot}}} \ h^{-1}$ bin, the mean and median values of our estimate are $\sim 40\%$ smaller than the average MAR values from the merger trees.
This systematic underestimate might be due to a statistical fluctuation because the sample only contains 50 clusters compared with the 2000 clusters of the less massive bin. In fact, a similar underestimate is observed when we compare the MAH from the merger trees with the Giocoli model in the right panel of Figure \ref{MAH_2R200}.
In contrast, for the $10^{14} \ {\rm{M_{\odot}}} \ h^{-1}$ mass bin, the mean and median MAR from our prescription recover the merger tree results within 20\% in the redshift range $z=[0,2]$. Our results are relevant because they show that our simple spherical infall prescription can in principle provide a method to estimate the average MAR of galaxy clusters from redshift surveys.
In future work, we will apply our prescription 
to synthetic redshift surveys of galaxy clusters to quantify the uncertainties and possible systematic errors of our procedure.

Clearly Figure \ref{spherical_infall} only compares the average MAR obtained from the merger trees of individual halos with the average MAR provided by our spherical infall technique. Our recipe was not conceived to completely capture all the features of the MAR derived by the complex merging process of individual halos.
Nevertheless, the average of the MAR of individual halos still is satisfactorily estimated by our recipe.

\begin{figure*}
\hbox{
 \epsfig{figure=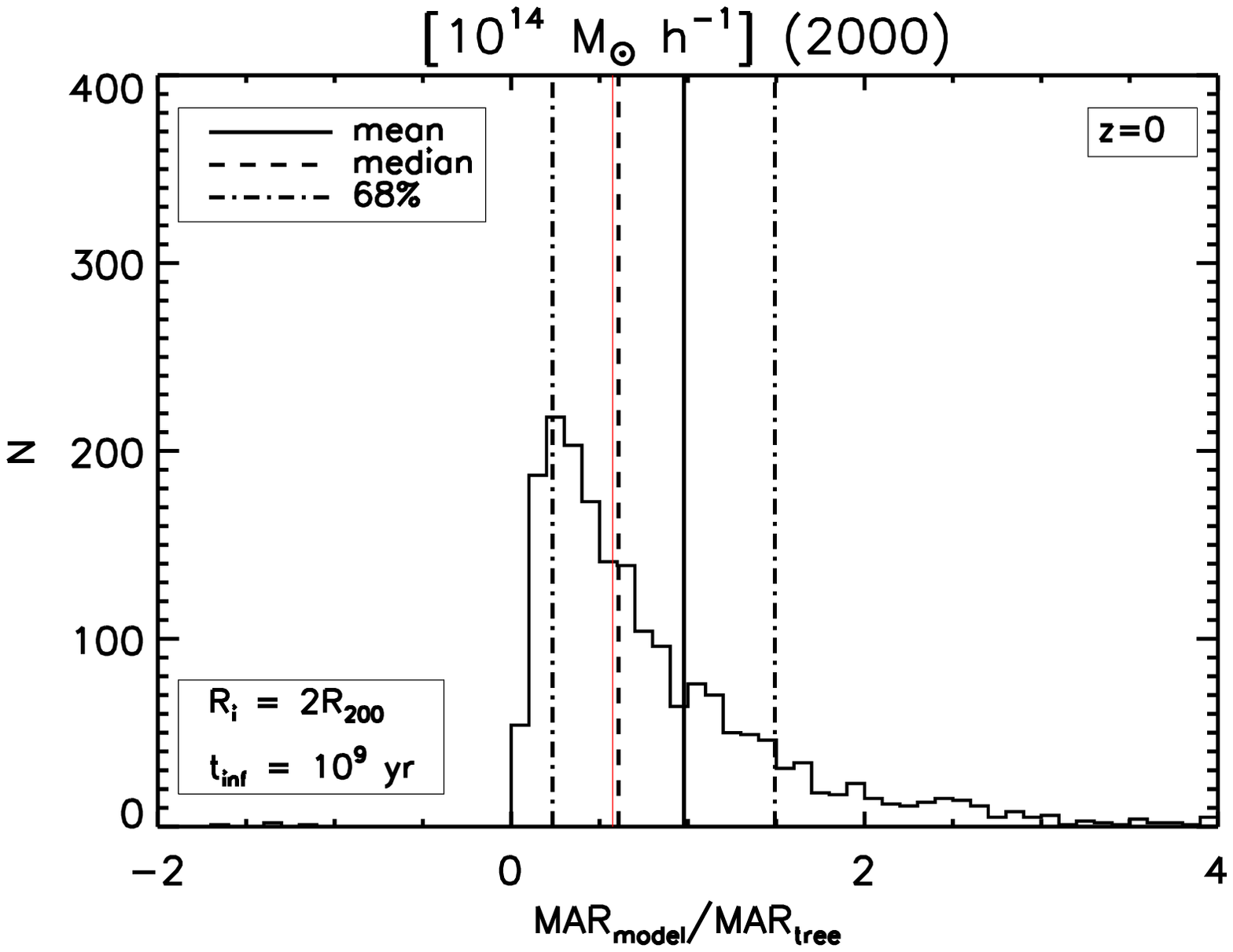,width=0.45\textwidth}
 \epsfig{figure=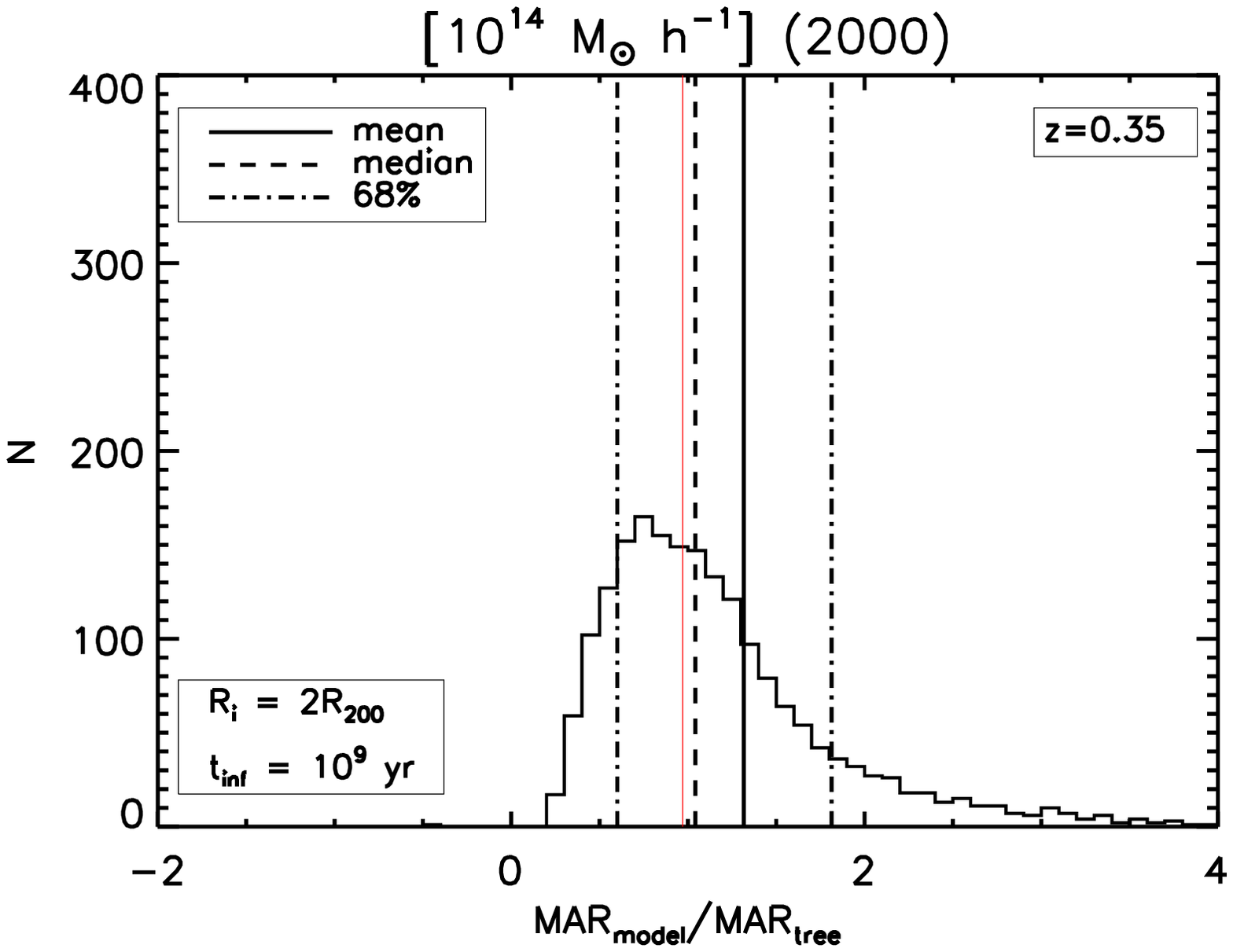,width=0.45\textwidth}
}
\hbox{
 \epsfig{figure=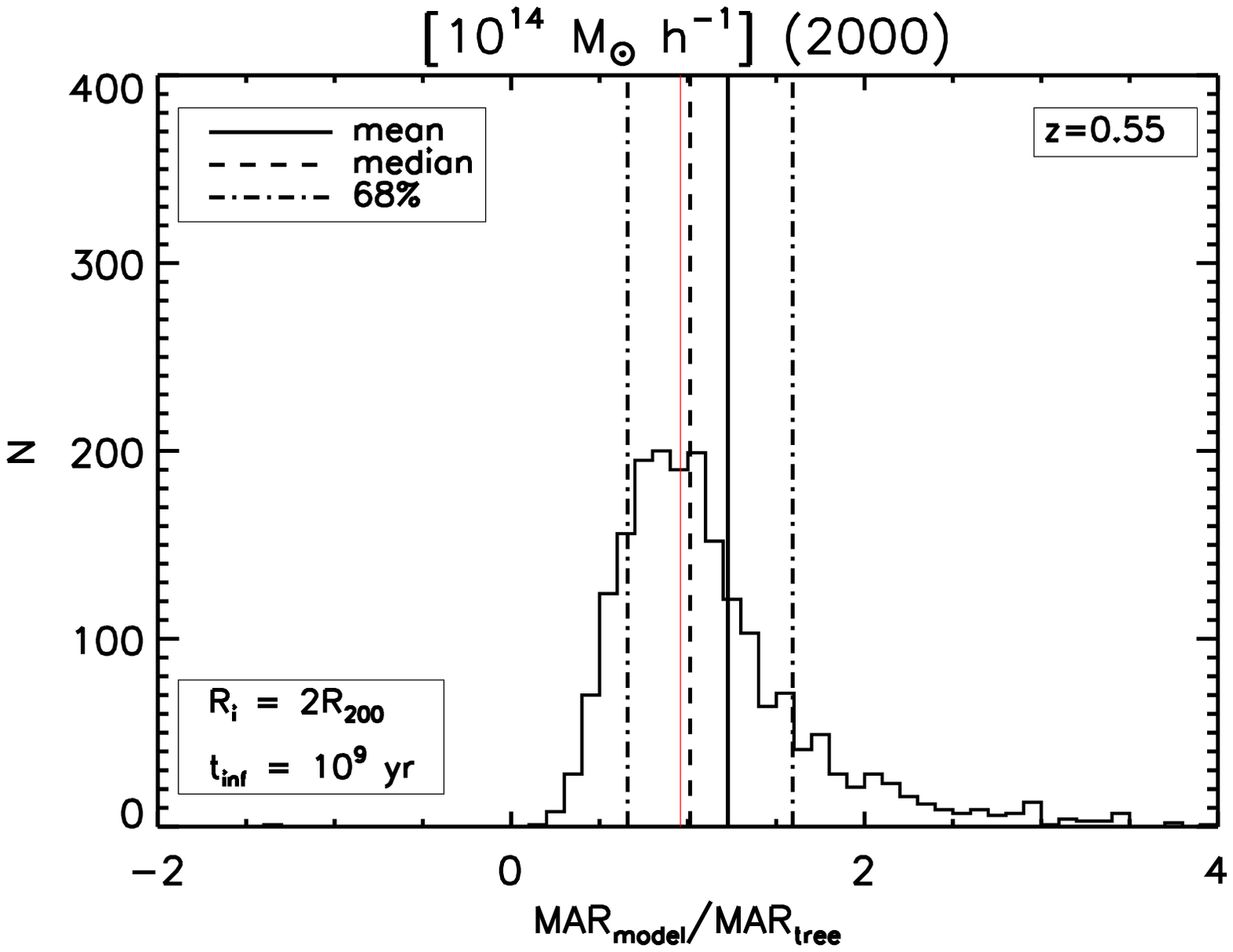,width=0.45\textwidth}
 \epsfig{figure=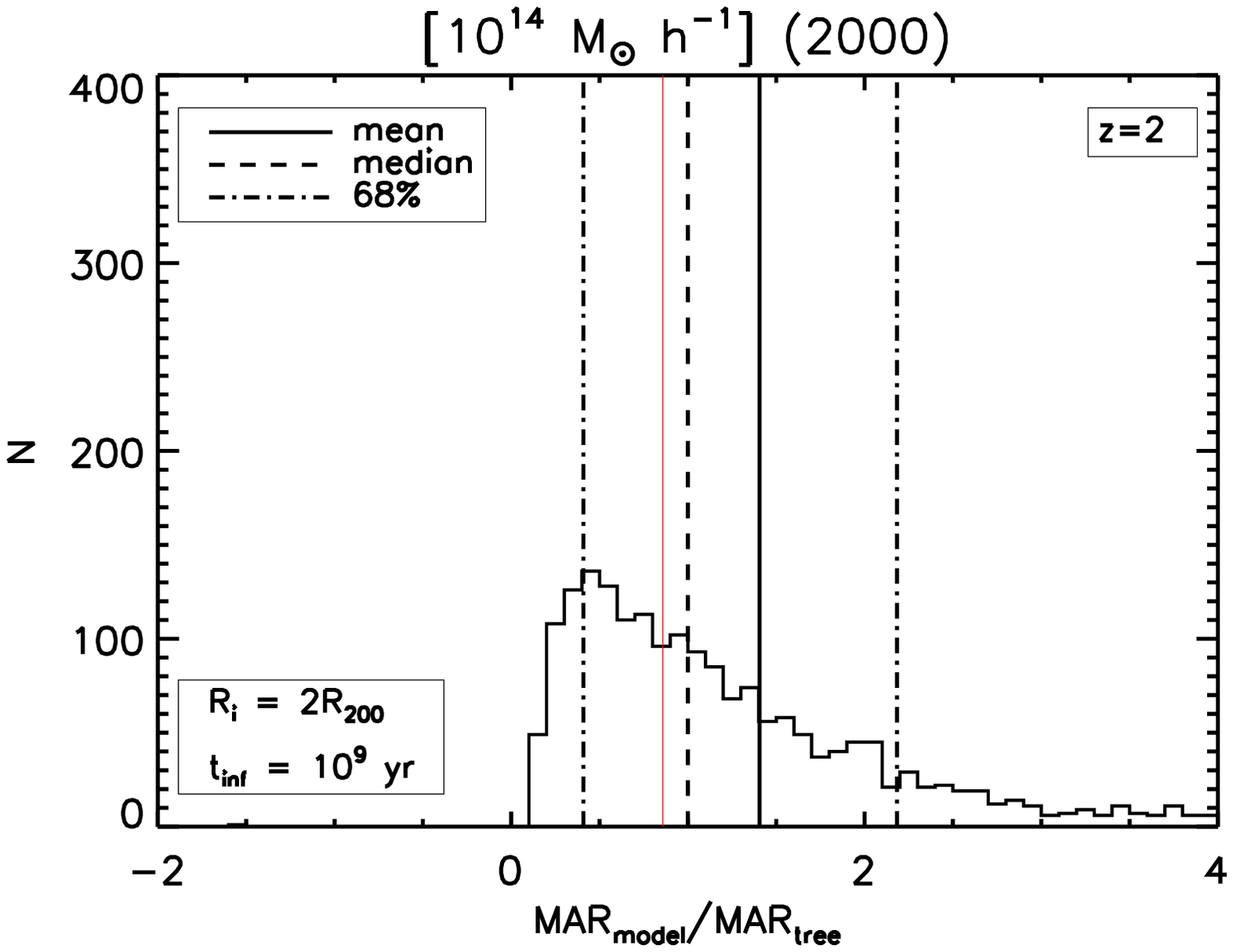,width=0.45\textwidth}
}
\hbox{
}
\caption{Histogram of the halo-by-halo ratio between the MAR from our spherical infall model and the MAR from the merger trees at $z=0$ (top-left panel), $z=0.35$ (top-right panel), $z=0.55$ (bottom-left panel) and $z=2$ (bottom-right panel) for the $10^{14} \ {\rm{M_{\odot}}} \ h^{-1}$ mass bin. The solid, dashed and dotted-dashed vertical lines mark the mean, median and $68 \%$ of the distribution, respectively. The red line marks the ratio between the average MAR from our model and the average MAR form the merger trees.}
\label{one-to-one_ratio}
\end{figure*}

Figure \ref{one-to-one_ratio} 
shows the distribution of the ratio between the MAR estimated with our recipe and the MAR derived from the merger tree for each individual halo of the $10^{14} \ {\rm{M_{\odot}}} \ h^{-1}$ mass bin at four different redshifts. 
The distribution has a tail toward large values. Remarkably the median value of this ratio is close to the ratio between the average MAR from our model and the MAR from the merger trees (red line). The mean clearly is larger because of the tail of high values. The $68 \%$ percentile ranges from $0.2$ to $2.2$. This result confirms that with our approach we can measure the mean MAR but not the MAR of the individual clusters, which is affected by lack of spherical symmetry and large variations of the infall velocity.

Figures \ref{velocity_profile_14}, \ref{velocity_profile_15}, and \ref{MAH_2R200}
show the intrinsic halo-by-halo scatter both in radial velocity and mass.  
Our choice to use the same $v_{\rm shell}$ for all clusters in a given mass bin at a given redshift implies that we neglect the scatter that originates the spread of the distribution shown in Figure \ref{one-to-one_ratio}. However, this choice keeps the model relatively simple and applicable to real clusters.
It is worth saying that the impact of the large-value tail is reduced if we take the ratio of the
averages of the MAR of each individual halo estimated with our prescription and with the merger trees, rather than the average of the ratio. The ratio of averages is shown with the red lines in Figure \ref{one-to-one_ratio} and it corresponds to the result shown in Figure \ref{spherical_infall}. The remarkable and encouraging result of our analysis is that the agreement shown in Figures \ref{spherical_infall} and \ref{one-to-one_ratio} is obtained without requiring any input information from the merger trees.

\section{Conclusions} \label{conclusions}

We investigate the feasibility of directly measuring the mean MAR of a sample of galaxy clusters from their mass profile. To measure the mean MAR we suggest a prescription based on the infall of a spherical shell of matter onto the halo, with constant acceleration and initial velocity 
derived from the average infall velocity of matter around dark matter halos in $N$-body simulations. Once we fix the scale $R_{i}$, that defines the halo radius at which the infall occurs, the initial velocity $v_{\rm shell}$ and the infall time $t_{\rm inf}$ our method only depends on the mass profile at large radii, beyond $R_{i}$.

We consider dark matter halos from the CoDECS set of $N$-body simulations and compare their MAR estimated with our prescription with the MAR estimated from the merger trees extracted from the simulations. We focus on two sets of halos with mass $M_{200}$  
around  $10^{14}$ and $10^{15} \ {\rm{M_{\odot}}} \ h^{-1}$ at $z=0$. 

We recover the mean and median MAR obtained from the merger trees without bias and with $20 \%$ accuracy in the redshift range $z=[0,2]$ for the $10^{14} \ {\rm{M_{\odot}}} \ h^{-1}$ mass bin. The accuracy is about $40 \%$, with a systematic underestimation of $\sim 40\%$, for the $10^{15} \ {\rm{M_{\odot}}} \ h^{-1}$ mass bin. This result is impressive given the simple assumptions of our prescription and the fact that no input parameter of the model is taken from the merger trees. Our result does show that measuring the mean MAR of a sample of real galaxy clusters is in principle possible.

A fundamental step to assess the feasibility of our approach is to apply the caustic technique to realistic
mock redshift surveys of galaxy clusters extracted from $N$-body simulations and quantify the accuracy of the estimated MAR. This investigation remains for further studies. Similarly the analysis of the dependence of the method parameters and its results on the cosmological model and on the theory of gravity remain 
to be investigated. Specifically, $v_{\rm shell}$ might turn out to vary substantially with the assumed cosmology and thus to be a crucial parameter 
of the method.

We might expect that the accuracy will be better than $20\%$ on the average MAR if we estimate the MAR of individual clusters in a given mass bin and estimate their average MAR; this approach would minimize the systematic errors due to projection effects which dominate the
estimate of the mass profile with the caustic technique. \cite{2013ApJ...767...15R} have already applied this approach to measure the total mass of clusters within their turnaround radius, the ultimate mass $M_{\rm ta}$. By combining $50$ CIRS clusters with $58$ HeCS clusters, they find $M_{\rm ta}/M_{200}=1.99 \pm 0.11$, a measure accurate to $5\%$ and in agreement with the $\Lambda$CDM prediction where $M_{\rm ta}/M_{200}$ has a log-normal distribution with a peak at mass ratio $2.2$ and dispersion $0.38$ \citep{2005MNRAS.363L..11B}.

Our measurements of the average MAR may provide an additional tool to discriminate among different cosmological models if deviations from the $\Lambda$CDM model generate measurable differences in the MAR. \\

We are deeply grateful to the anonymous referee whose very accurate comments and suggestions urged us to substantially revise our analysis which led to more solid and valuable results.
We also sincerely thank Fabio Fontanot for providing us with the merger trees of the H-CoDECS simulation. We thank Aaron Ludlow, Margherita Ghezzi, Giulio Falcioni, and Andrea Vittino for useful discussions. 
C. D. B., A. L. S., and A. D. acknowledge partial support from the grant Progetti di Ateneo/CSP TO$\_$Call2$\_$2012$\_$0011 ``Marco Polo" of the University of Torino, the INFN grant InDark, and the grant PRIN$\_$2012 ``Fisica Astroparticellare Teorica" of the Italian Ministry of University and Research. A. L. S. also acknowledges the support of Dipartimento di Fisica, University of Torino, where most of this project was carried out. 
C. G.'s research is part of the project GLENCO, funded under the European Seventh Framework Programme, Ideas, Grant Agreement n. 259349. C. G. also thanks CNES for financial support.
M. B. is supported by the Marie Curie Intra-European Fellowship ``SIDUN" within the 7th Framework Programme of the European Commission.
This work was partially supported by grants from R\'egion Ile-de-France.

\appendix

In this appendix, we investigate the MAH of $M_{200}$ of dark matter halos: we compare the merger tree results with two known fitting formulae and a theoretical model.
We calculate the mean and median MAH for the objects in four mass bins (Table \ref{mass_bins}). The two most massive bins are the same used in the main body of the paper. The four bins are centered on $M_{200}= 10^{15}, 10^{14}, 10^{13}, 5 \times 10^{12} \ {\rm{M_{\odot}}} \ h^{-1}$. Each bin contains $2000$ objects at $z=0$ with the exception of the largest-mass bin whose sample is limited to $50$ objects. The lowest-mass bin is centered on $M_{200} = 5 \times 10^{12} \ {\rm{M_{\odot}}} \ h^{-1}$, because $M_{200} = 10^{12} \ {\rm{M_{\odot}}} \ h^{-1}$ is below the resolution limit of about 100 particles per subhalo set by SUBFIND. Table \ref{mass_bins} also lists the $90 \%$ percentile mass range of each mass bin.

\begin{table*}
\begin{center}
\caption{Mean, Median, Standard Deviation, and Percentile Ranges of $M_{200}$ and Number of Halos in Each Mass Bin at $z=0$}
\begin{tabular}{lccccc}
\hline
\hline
\\
Mean & $\sigma$ & Median & $68\%$ & $90\%$ & Number of Halos \\
\\
\hline
\\
\multicolumn{5}{c}{$M_{200} \ [{\rm{M_{\odot}}} \ h^{-1}]$} & \\
\\
\cline{1-5}
\\
$1.04 \times 10^{15}$ & $1.26 \times 10^{14}$ & $1.00 \times 10^{15}$ & $(0.91-1.19) \times 10^{15}$ & $(0.88-1.24) \times 10^{15}$ & $50$\\
\\
\hline
\\
$1.00 \times 10^{14}$ & $2.90 \times 10^{12}$ & $1.00 \times 10^{14}$ & $(0.97-1.04) \times 10^{14}$ & $(0.96-1.05) \times 10^{14}$ & $2000$ \\
\\
\hline
\\
$1.00 \times 10^{13}$ & $2.36 \times 10^{10}$ & $1.00 \times 10^{13}$ & $(1.00-1.00) \times 10^{13}$ & $(1.00-1.00) \times 10^{13}$ & $2000$ \\
\\
\hline
\\
$5.00 \times 10^{12}$ & $6.04 \times 10^{9}$ \ \ & $5.01 \times 10^{12}$ & $(5.00-5.01) \times 10^{12}$ & $(5.00-5.01) \times 10^{12}$ & $2000$ \\
\\
\hline
\end{tabular}
\label{mass_bins}
\end{center}
\end{table*}

Figure \ref{MAH_comparison} shows the MAHs of the four mass bins. We limit our study to the low redshift range $z=[0,2]$ because we are interested in the observational relevance of our analysis. As we can see in Figure \ref{MAH_comparison}, for the two largest-mass bins the mean MAH agrees with the median MAH within $20 \%$. In the two smallest-mass bins the difference between the mean and the median MAH is never larger than 5\%. In all four cases the standard deviation and the $68\%$ percentiles are comparable. The results from the largest-mass bin are noisier because of the low-number statistics. The number of objects $N_{\rm hal}$ at each $z$ decreases with increasing $z$, due to the resolution limit: not all the objects selected at $z=0$ have merger trees that reach $z=2$. Indeed, the decrease is larger for less massive objects which are already closer to the mass resolution limit at $z=0$.

\subsection{A.1. Fitting Formulae} \label{fit_MAH}

\begin{figure*}
\hbox{
}
\hbox{
}
\hbox{
}
\hbox{
 \epsfig{figure=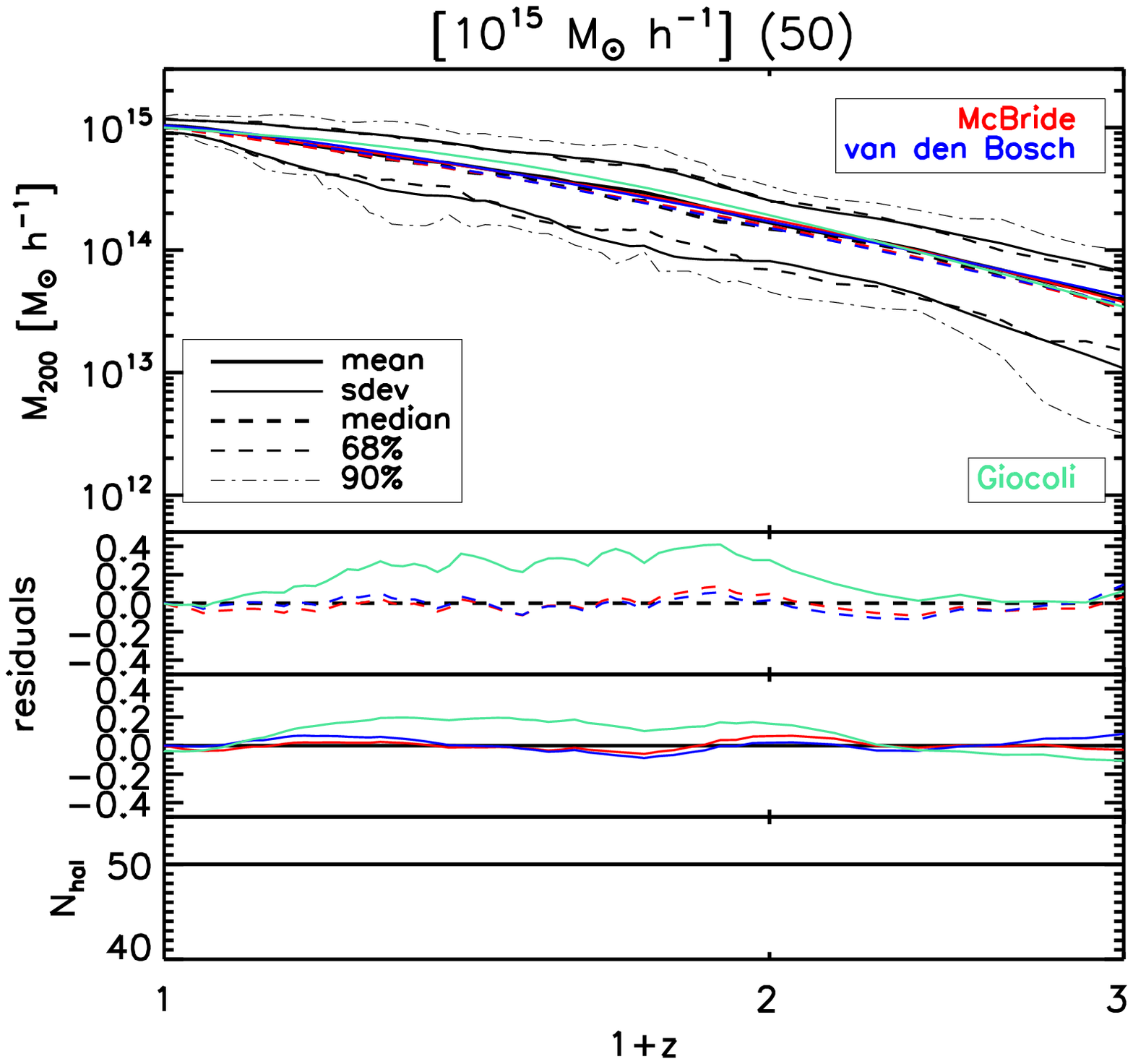,width=0.45\textwidth}
 \epsfig{figure=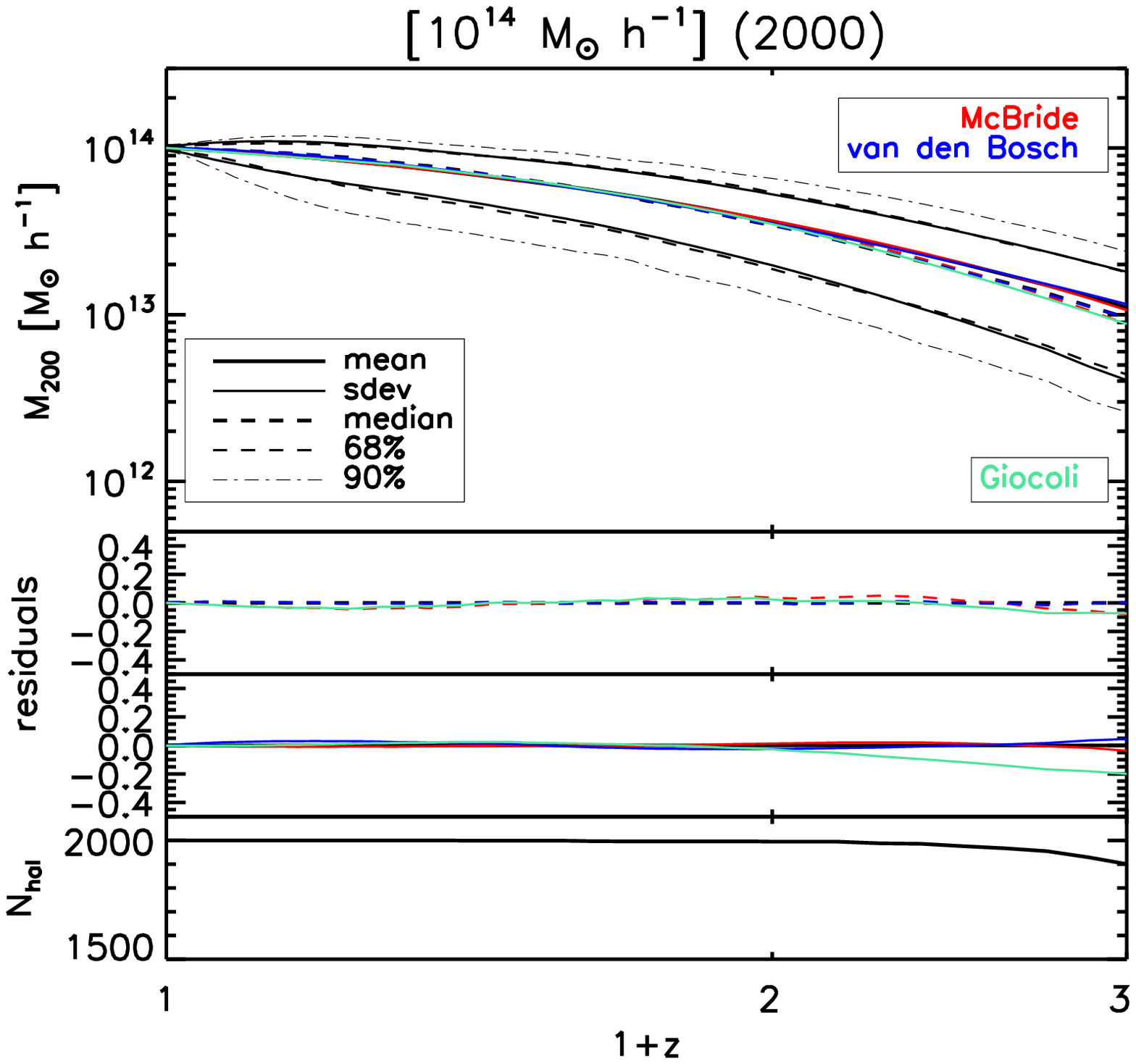,width=0.45\textwidth}
}
\hbox{
}
\hbox{
}
\hbox{
}
\hbox{
}
\hbox{
}
\hbox{
 \epsfig{figure=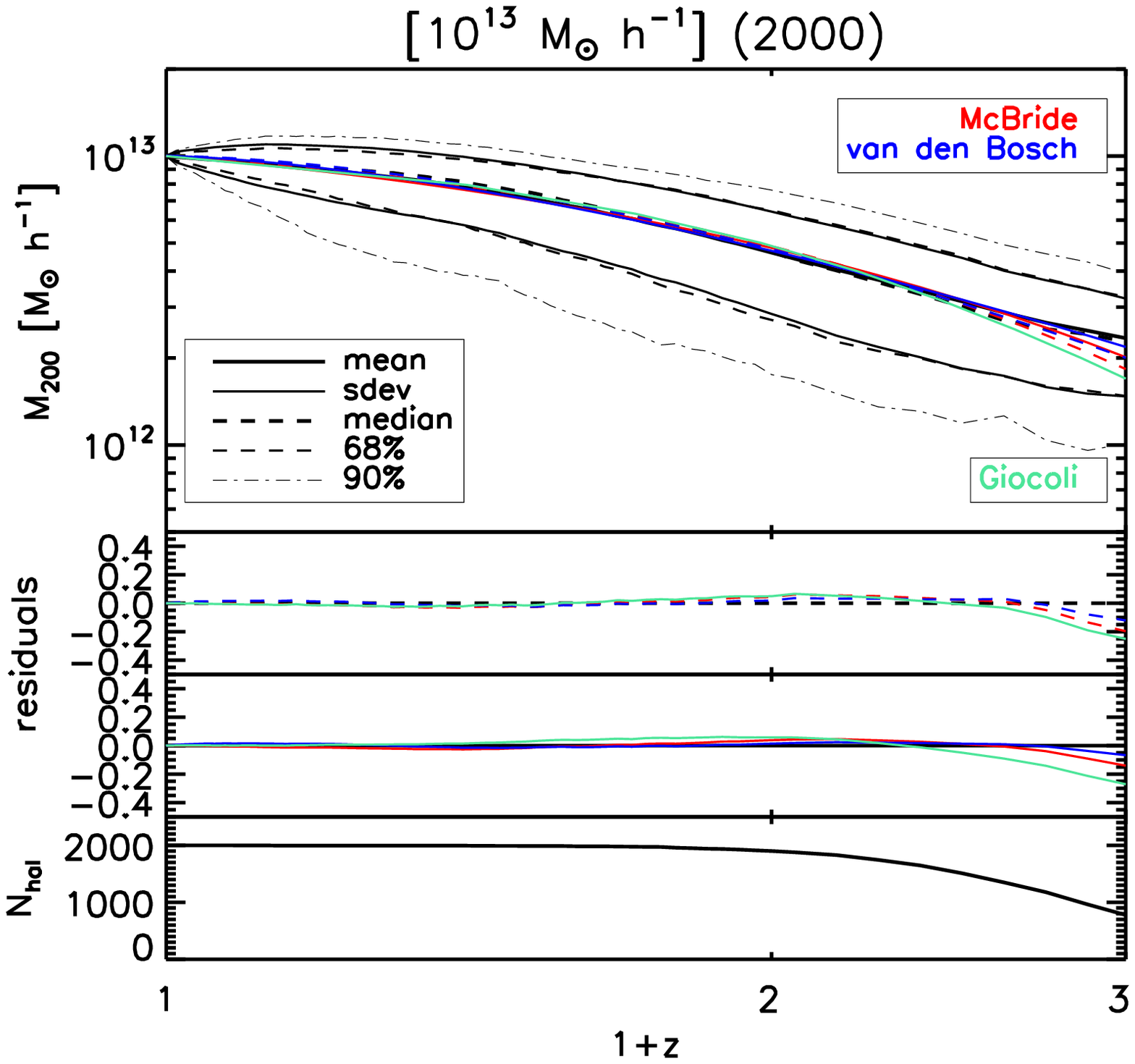,width=0.45\textwidth}
 \epsfig{figure=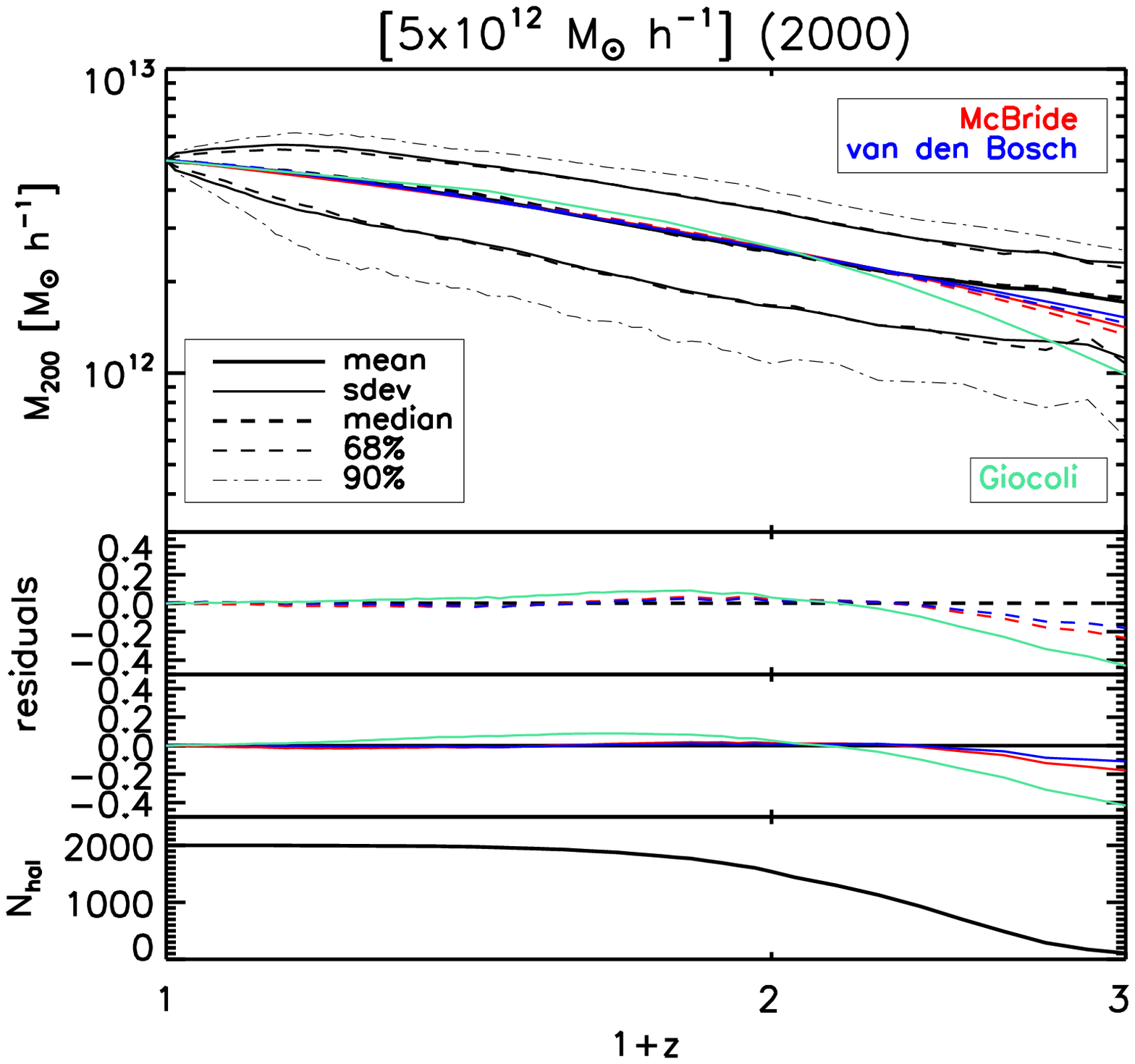,width=0.45\textwidth}
}
\hbox{
}
\hbox{
}
\hbox{
}
\caption{$M_{200}(z)$ in the four mass bins as indicated above each panel. The thick-solid and thick-dashed lines indicate the mean and median MAH, respectively. The thin-solid and thin-dashed lines indicate the standard deviation and the $68\%$ of the distribution, respectively. The thin-dotted-dashed line indicates the $90\%$ of the distribution. The best fits and theoretical model are overplotted: the curves from Equation (\ref{van_den_Bosch_mass}) are in blue, those from Equation (\ref{McBride_mass}) are in red, and those fro the Giocoli model are in green. Residuals from the median and the mean as well as the total number of objects at each redshift are shown in the insets at the bottom of each panel.}
\label{MAH_comparison}
\end{figure*}

Different fitting formulae for the MAH shown in Figure \ref{MAH_comparison} exist in the literature. We focus on two of them. By using the extended Press-Schechter (EPS) formalism \citep{1991ApJ...379..440B,1993MNRAS.262..627L}, \cite{2002MNRAS.331...98V} proposed 

\begin{equation}
M(z)= M_{0} \exp \left\{ \ln(1/2) \left[ \frac{\ln(1+z)}{\ln(1+z_{f})} \right]^{\nu} \right\} \ , \tag{4}
\label{van_den_Bosch_mass}
\end{equation}

\noindent where $z_{f}$ and $\nu$ are free parameters. The redshift formation $z_{f}$ indicates the redshift when the halo has accreted half of its final mass. We fit both the median and the mean MAH with the equation above by assuming Poisson errors weighted by the number of halos in each redshift bin. We quantify the deviation from this analytic description of the MAH with the rms of the fit 

\begin{equation}
(r.m.s.)^{2}= \frac{1}{N} \sum_{N} \frac{\left[ M(z_{i}) - M_{\rm model}(z_{i}) \right]^{2}}{M(z_{i})^{2}} \ . \tag{5}
\label{rms}
\end{equation}

We list the best-fit parameters of Equation (\ref{van_den_Bosch_mass}) along with the rms of the fits in Table \ref{MAH_best_fit}. As expected in hierarchical clustering scenarios, the value of the best-fit parameter $z_{f}$ increases with decreasing mass because more massive objects tend to form later than less massive ones.

\begin{table*}
\begin{center}
\caption{Best-fit Parameters of the Formulae by Van Den Bosch, Equation  (\ref{van_den_Bosch_mass}), and McBride, Equation (\ref{McBride_mass})}
\begin{tabular}{lcccccccc}
\hline
\hline
\\
& \multicolumn{4}{c}{Median} & \multicolumn{4}{c}{Mean} \\
\\
\hline
\\
& \multicolumn{8}{c}{van den Bosch} \\
\\
\cline{2-9}
\\
$M_{200} [10^{14} {\rm{M_{\odot}}} \ h^{-1}]$ & $z_{f}$ & $\nu$ & & rms & $z_{f}$ & $\nu$ & & rms \\
\\
\hline
\\
$10$ & $0.358 \pm 0.011$ & $1.227 \pm 0.035$ & & $0.044$ & $0.381 \pm 0.012$ & $1.252 \pm 0.037$ & & $0.043$ \\
$1$ & $0.711 \pm 0.003$ & $1.702 \pm 0.012$ & & $0.006$ & $0.709 \pm 0.003$ & $1.582 \pm 0.012$ & & $0.022$ \\
$0.1$ & $0.944 \pm 0.005$ & $1.675 \pm 0.019$ & & $0.025$ & $0.925 \pm 0.005$ & $1.519 \pm 0.018$ & & $0.015$ \\
$0.05$ & $1.033 \pm 0.010$ & $1.317 \pm 0.024$ & & $0.032$ & $1.036 \pm 0.011$ & $1.242 \pm 0.022$ & & $0.024$ \\
\\
\hline
\\
& \multicolumn{8}{c}{McBride} \\
\\
\cline{2-9}
\\
$M_{200} [10^{14} {\rm{M_{\odot}}} \ h^{-1}]$ & $\beta$ & $\gamma$ & $\beta -\gamma$ & rms & $\beta$ & $\gamma$ & $\beta -\gamma$ & rms \\ 
\\
\hline
\\
$10$ & $-1.000 \pm 0.270$ & $1.150 \pm 0.180$ & $-2.160$ & $0.048$ & $-0.690 \pm 0.270$ & $1.280 \pm 0.180$ & $-1.970$ & $0.028$ \\
$1$ & $+1.283 \pm 0.044$ & $1.920 \pm 0.029$ & $-0.637$ & $0.030$ & $+0.812 \pm 0.043$ & $1.566 \pm 0.028$ & $-0.754$ & $0.010$ \\
$0.1$ & $+0.898 \pm 0.051$ & $1.342 \pm 0.035$ & $-0.444$ & $0.039$ & $+0.480 \pm 0.051$ & $1.065 \pm 0.035$ & $-0.585$ & $0.029$ \\
$0.05$ & $-0.005 \pm 0.070$ & $0.658 \pm 0.051$ & $-0.664$ & $0.053$ & $-0.217 \pm 0.069$ & $0.513 \pm 0.050$ & $-0.730$ & $0.037$ \\
\\
\hline
\end{tabular}
\label{MAH_best_fit}
\end{center}
\end{table*}

The second formula we considered was first proposed by \cite{2004ApJ...607..125T} and widely studied by \cite{2009MNRAS.398.1858M}: 

\begin{equation}
M(z) = M_{0} (1+z)^{\beta} e^{-\gamma z} \ , \tag{6}
\label{McBride_mass}
\end{equation}

\noindent where $\beta$ and $\gamma (\ge 0)$ are free parameters. This formula represents an exponential growth with a redshift-dependent correction. It is a revision of the simple one-parameter exponential form $M(z)=M_{0}e^{-\alpha z}$ \citep[][]{2002ApJ...568...52W}, where $\alpha = \ln(2) / z_{f}$. By using the EPS formalism, \cite{2015MNRAS.450.1514C} showed that in a $\Lambda$CDM model the exponential growth is a good description of the MAH at high $z$, while the power-law behavior at low $z$ is necessary because the accelerated expansion of the universe slows down the accretion. For this reason Equation (\ref{McBride_mass}) appears to be a general description of the MAH of dark matter halos in a $\Lambda$CDM model independently of the halo mass. The parameters $\beta$ and $\gamma$ are related to the power spectrum \citep{2015MNRAS.450.1514C}. The value of $\beta-\gamma$ is a parameter describing the mass growth rate at low redshift. We use Equation (\ref{McBride_mass}) to perform a fit with Poisson errors weighted by the number of halos in each redshift bin and evaluate the rms as in Equation (\ref{rms}). We list the best-fit parameters of Equation (\ref{McBride_mass}) and the rms of the fits in Table \ref{MAH_best_fit}.

\subsection{A.2. Comparison with a Theoretical Model} \label{theoretical_MAH}

By following and generalizing the formalism of \cite{1993MNRAS.262..627L} and \cite{1999MNRAS.303..685N}, \cite{2012MNRAS.422..185G} introduced a new theoretical model to describe the MAH of dark matter halos. This simple model, which enables the derivation of a generalized redshift formation distribution, has already been applied to the CoDECS simulations in \cite{2013MNRAS.434.2982G}. 

Here we summarize the relevant definitions and refer to \cite{2012MNRAS.422..185G} for further details.
The model defines the redshift formation $z_{f}$ of a halo of a given mass $M_{0}$ at a given redshift $z_{0}$ as the redshift when the object has accreted a fraction $f$ of its final mass $M_{0}$, for any fraction $0 < f < 1$. 
The variance of the linear fluctuation field is 

\begin{equation}
S(M) = \frac{1}{2\pi^2} \int_{0}^{\infty} W^2 (kR) P_{\rm lin} (k) k^2 dk \ , \tag{7}
\end{equation}

\noindent where $W(kR)$ is a top-hat window function of scale $R=(3M/4\pi \rho_m)^{1/3}$, $\rho_m$ is the comoving background density of the universe, and $P_{\rm lin} (k)$ is the linear power spectrum. 

The initial threshold overdensity for spherical collapse is

\begin{equation}
\delta_{c}(z) = \frac{\delta_{c,\rm DM}}{D_{+}(z)} \ , \tag{8}
\end{equation}

\noindent where $\delta_{c,\rm DM}$ is the linear overdensity at redshift $z$ and $D_{+}(z)$ is the growth factor normalized to unity at $z=0$.

The cumulative formation redshift distribution, in terms of the scaled variable

\begin{equation}
w_{f} = \frac{\delta_c(z_{f}) - \delta_c(z_{0})}{\sqrt{S(fM) - S(M)}} \ , \tag{9}
\end{equation}

\noindent is given by

\begin{equation}
P(>w_{f}) = \frac{\alpha_{f}}{e^{w_{f}^2 /2} + \alpha_{f} -1} \ . \tag{10}
\label{Giocoli_probability}
\end{equation}

The model has a single free parameter $\alpha_{f}$ which depends on the fraction $f$ of the final mass that is used to define the formation redshift $z_{f}$. For the same set of simulations we use here, \cite{2013MNRAS.434.2982G} find

\begin{equation}
\alpha_{f} = \frac{1.365}{f^{0.65}} {\rm{e}}^{-2f^3} \ .  \tag{11}
\end{equation}

Since Equation (\ref{Giocoli_probability}) can be inverted, it is possible to evaluate the median redshift $z_{f}$ when a halo accretes a fraction $f$ of its final mass $M$ with the relation 

\begin{equation}
 \delta_{c}(z_{f}) = \delta_{c}(z_{0}) + \tilde{w_{f}} \sqrt{S(fM) - S(M)} \ , \tag{12}
\label{Giocoli_zf}
\end{equation}

\noindent where 

\begin{equation}
\tilde{w_{f}} = \sqrt{2 \ln(\alpha_{f} +1)} \tag{13}
\end{equation}

\noindent is the median value of $w_{f}$ defined by the usual relation $P(>\tilde w_{f}) = 1/2$. Equation (\ref{Giocoli_zf}) can be translated into a MAH for a given final mass $M_{0}$. 

We compare the simulation results and the Giocoli model in Figure \ref{MAH_comparison}. The model of \cite{2012MNRAS.422..185G} is built by evaluating the median redshift at which the halo has accreted a fixed fraction of its final mass whereas we evaluate the mean and the median $M_{200}$ for all the objects in a given mass bin at each redshift.
We list the rms defined in Equation (\ref{rms}) in Table \ref{MAH_Giocoli_rms}. 
In general, the global agreement between the theoretical model of \cite{2012MNRAS.422..185G} and the simulation in each mass bin is similar for both the mean and the median, as it can be seen from the rms values in Table \ref{MAH_Giocoli_rms}. In the largest-mass bin the model overestimates the MAH obtained from the simulation, while in the other three bins the agreement is within a few percent up to $z \sim 1$. Toward higher $z$, the model starts underestimating the simulation MAH. This discrepancy is more pronounced and appears at decreasing redshifts for decreasing halo mass. This behavior originates from the mass resolution and the consequent decrease of the number of halos $N_{\rm hal}$ at a given redshift.

\begin{table}
\begin{center}
\caption{MAH rms for the Giocoli model}
\begin{tabular}{lcc}
\hline
\hline
\\
& Median & Mean \\
\\
\hline
\\
$M_{200} [10^{14} {\rm{M_{\odot}}} \ h^{-1}]$ & \multicolumn{2}{c}{rms} \\ 
\\
\cline{2-3}
\\
$10$ & $0.144$ & $0.136$ \\
$1$ & $0.058$ & $0.052$ \\
$0.1$ & $0.059$ & $0.058$ \\
$0.05$ & $0.101$ & $0.103$ \\
\\
\hline
\end{tabular}
\label{MAH_Giocoli_rms}
\end{center}
\end{table}

\bibliographystyle{apj}
\bibliography{cristiano}

\end{document}